\RequirePackage{lineno}%DELETE FOR PLAIN VERSION
\documentclass[superscriptaddress,aps,preprint,amsmath,amssymb,floatfix]{revtex4-1}

\usepackage{graphicx,morefloats}
\usepackage{subfigure} % referring to Figure a and b separately
\usepackage{color,soul}
\usepackage{bm}
\usepackage{amsmath}
\usepackage{amssymb}
\usepackage{times}
\usepackage{hyperref}
\usepackage{dsfont}
\usepackage{bbm}
\usepackage{xcolor}
\usepackage{gensymb}

\newcommand{\be}[0]{\begin{equation}}
\newcommand{\ee}[0]{\end{equation}}
\newcommand{\ba}[0]{\begin{eqnarray}}
\newcommand{\ea}[0]{\end{eqnarray}}
\newcommand{\mx}[0]{\begin{pmatrix}}
\newcommand{\ex}[0]{\end{pmatrix}}

\usepackage{hyperref}
\makeatletter
\newcommand*{\rom}[1]{\expandafter\@slowromancap\romannumeral #1@}
\makeatother
\newcommand{\sgn}{\operatorname{sgn}}

\begin{document}
%\linenumbers
%%%%%% USE: Ref.\citenum{xxx} INSTEAD OF Ref.\,\onlinecite{xxx} %%%%%

\hyphenation{va-ni-sh-ing}

\begin{center}

\thispagestyle{empty}

{\large\bf 
Quantum Criticality Enabled by Intertwined Degrees of Freedom
}
\\%[0.6cm]
[0.3cm]

Chia-Chuan\ Liu$^{1,2}$,
Silke\  Paschen$^{3,1}$, and Qimiao Si$^{1,\ast}$
\\[0.3cm]

$^1$
Department of Physics and Astronomy, Rice Center for Quantum Materials,
Rice University, Houston, Texas 77005, USA
\\[-0.cm]

$^2$
Département de Physique,  Université de Montréal, Québec,  H3C 3J7, Canada
\\[-0.cm]

$^3$Institute of Solid State Physics, Vienna University of Technology, 1040 Vienna, Austria

\end{center}

%\vspace{0.3cm}
\vspace{0.16cm}

{\bf 
Strange metals appear in a wide range of correlated materials. 
Electronic localization-delocalization and the expected loss of quasiparticles 
characterize beyond-Landau metallic quantum critical points and the associated strange metals.
Typical settings involve local spins. Systems that contain entwined degrees of freedom
offer new platforms to realize novel forms of quantum criticality. 
Here, we study the fate of an SU(4) spin-orbital Kondo state in a multipolar Bose-Fermi Kondo model, 
which provides an effective description of a multipolar Kondo lattice, using a renormalization-group method.
We show that at zero temperature a generic trajectory in the model's parameter space 
contains two quantum critical points, which are associated with the destruction of 
Kondo entanglement in the orbital and spin channels respectively. 
Our asymptotically exact results reveal an overall phase diagram, provide the theoretical basis
to understand puzzling recent experiments of a multipolar heavy fermion metal,
and point to a means of designing new forms of quantum criticality and strange metallicity
 in a variety of strongly correlated systems.
}
\vspace{0.6cm}

\noindent
{\it {\bf Significance Statement:~~}
Melting ice illustrates how phase transitions occur by varying temperature.  
Quantum phase transitions appear at absolute zero temperature, when the extent to which Heisenberg's 
uncertainty principle affects matter is tuned through a control parameter. For a continue transition, 
quantum criticality arises and influences the physics over a wide parameter range at finite temperatures. 
In quantum materials, the microscopic agent for quantum criticality is usually spin. 
Here we show that intertwining spins, orbitals and other degrees of freedom provides a means 
to design novel forms of quantum criticality. Our work provides the understanding of puzzling recent experiments 
in a spin-orbital-entwined heavy fermion metal, and promises to realize new types 
of quantum criticality in a variety of strongly correlated metals.
}
\vspace{0.6cm}

%\noindent E-mails: qmsi@rice.edu

\newpage
%%%%%%%%%%%%%%%%%%%%%%%%%%%%%%%%%%%%%%%%%%%%%%%%
Simple metals such as copper and aluminum are well described in terms of weakly correlated itinerant electrons.
In a wide range of strongly correlated metals, the electrons' Coulomb repulsion
is comparable to or larger than their bandwidth \cite{Kei2017,PaschenSi-2020}. 
The strong correlations are expected to cause
 a loss of Landau quasiparticles and the associated strange metallicity \cite{Hu-qcm2022.2,Phillips2022}.
Correlations turn certain bare electrons into effective
local degrees of freedom 
in the building blocks 
for the low-energy physics.
A prototypical case is the 
heavy fermion metals, which feature 
a wide variety of quantum phases
\cite{Stewart2001,Coleman2005,Kirchner2020}.
Here, local spins 
are associated with the correlated
4$f$-electrons. Their entanglement with the background conduction
electrons gives rise to the spin-isotropic [SU(2)-symmetric] Kondo effect \cite{Hewson-book}.
The destruction of the Kondo effect corresponds to a localization of the 4$f$-electrons,
is expected to cause a
 loss of quasiparticles, 
and represents a prototype 
mechanism for strange 
metallicity and 
beyond-Landau quantum critical points 
(QCPs) \cite{Qimiao-Nature,Coleman-JPCM,Senthil-PRB,Schroder-Nature,Paschen-Nature,Shishido-JPSJ,Park-Nature,Prochaska-Science,Nguyen-NatComm2021}.

The notion that local correlation effects drive new forms of quantum fluctuations 
raises the possibility of designing new types of quantum criticality by controlling local degrees of freedom.
For the Kondo effect {\it per se}, various kinds of local degrees of freedom have led to a variety of Kondo 
states 
relevant to multipolar heavy fermion 
metals \cite{Patri2020, Patri2020prx, HsinHua2018,VanDyke2019,Zhang2018,Bolech2005,Bolech2002,Cox1998},
multi-orbital iron-based compounds \cite{Si2016,Aron2015,Walter2020,Ong2012},
synthetic systems 
such as ultracold atoms \cite{Nishida2013}
and mesoscopic devices
\cite{LeHur2003,Goldhaber-Gordon2007,Mitchell2020},
and other correlated settings \cite{horvat2016,Rau2014,Coleman1995,Affleck1991}.
Recent experiments \cite{Ye2021.x,Ekahana2022.x} have motivated the idea \cite{HaoyuHu2022.x,LeiChen2022.x}
that, through 
molecular orbitals 
(of limited spatial extent),
 Kondo effects develop as a proper description of the low-energy physics
 even for $d$-electron-based flat band systems.
 Meanwhile, in twisted graphene structures, there have been proposals for their understanding in terms of 
 Kondo effects that are 
 associated with the 
 degrees of freedom of moir\'{e} unit cells 
 \cite{Ramires2021,Song2022,Guerci2022.x}.
In these systems, different kinds of crystalline symmetries or stacking/twisting in different types of flat bands 
can yield various forms of local degrees of freedom.
In heavy fermion metals, the nature of the local degrees of freedom is controlled by
 the cooperation of strong correlations, large spin-orbit coupling and crystalline symmetry.
 Indeed, 
there is 
a growing list of heavy fermion metals in which the role
of multipolar degrees of freedom has been explored for their quantum criticality \cite{PaschenSi-2020,Paschen2014}.
These include Pr(TM)$_2$Al$_{20}$ (TM = Ti, V), which have non-magnetic doublets in the ground-state manifold
\cite{Shimura2015,Sakai2011},
 PrOs$_4$Sb$_{12}$, which involves field-induced local quadrupolar moments
 \cite{McCollam2013,Bauer2002},
 and 
 YbRu$_2$Ge$_2$, which hosts quasi-degenerate spin and higher-rank moments \cite{Rosenberg2019,Jeevan2006}.
 
The hope of advancing this design principle for new types of quantum criticality is in particular 
triggered by 
recent experimental 
studies \cite{Martelli2019}
on a heavy fermion compound Ce$_3$Pd$_{20}$Si$_6$
 (Ref.\citenum{Cus2012})
as  a function of a non-thermal control parameter (magnetic field).
Surprisingly, the experimental results show
 {\it two stages} of Kondo-destruction quantum criticality [see
the Supplementary Information (SI), Sec.\,A].
In this system,  the $4f$ electrons form a total angular momentum $J=5/2$ state
whose six-fold degeneracy
is further split as dictated by the point-group symmetry \cite{Shi1997}. 
What lies in the ground-state manifold is the $\Gamma_8$ quartet \cite{Lor2012},
which can be represented in the pseudo-spin $\vec{\sigma}$ and pseudo-orbital $\vec{\tau}$ bases (see the SI, Sec.\,A).
The competition between the Kondo entanglement in the $\Gamma_8$-manifold and
the associated RKKY interactions may therefore be responsible for this sequential Kondo destruction.

\begin{figure}[t]
%\begin{center}
%\includegraphics[width=0.9\columnwidth]{fig_latt.pdf}
%\vskip 0.15 in
%\subfigure[]{
\includegraphics
[scale=0.46]
%[scale=1.38]
%{BFKbath.png}
{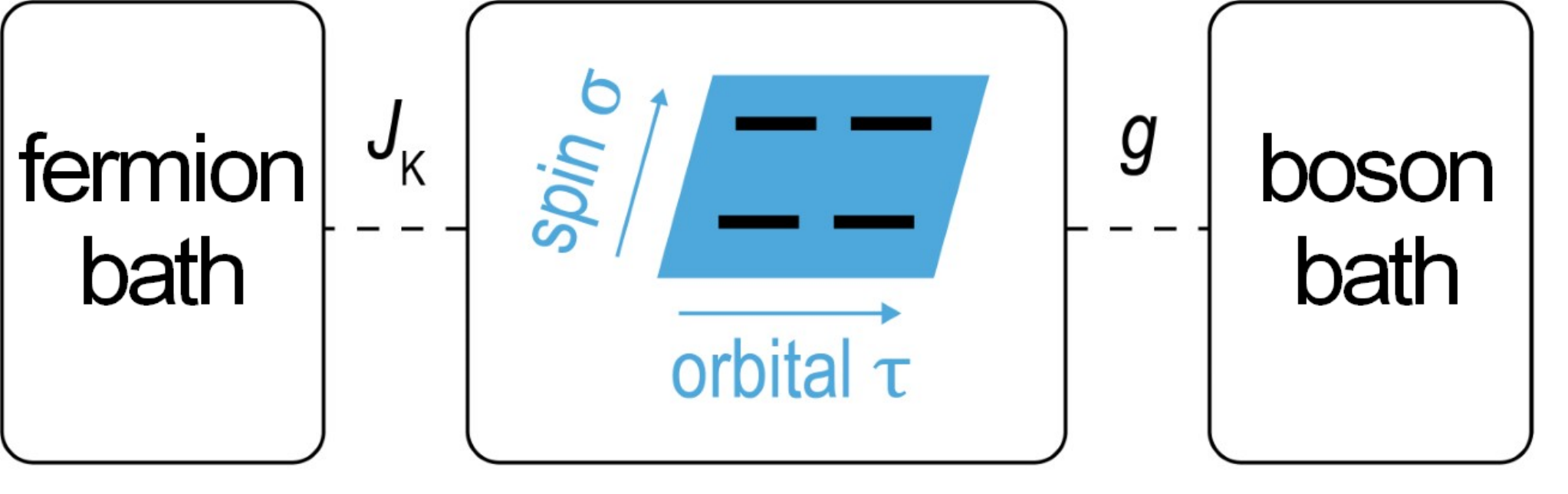}
%\end{center}
\vskip 0.05 in
% -0.15 in
\caption{ 
{\bf Illustration of the model.}
The multipolar Bose-Fermi Kondo model (Eq.\,\ref{SU4RGmodel})
describes entwined local spin-orbital degrees of freedom that are 
coupled to a bosonic and fermionic bath.
 }
 \label{fig:su4bfk2}
\end{figure}

The striking experimental observations 
motivate a well-defined theoretical
question: what is the generic type of QCPs that result from this type of competition?
The minimal prototype model of interest
is the spin-orbital entwined multipolar Bose-Fermi Kondo 
model (BFK) \cite{Martelli2019,Hu-edmft2022.3}, as
illustrated in Fig.\,\ref{fig:su4bfk2}.
 It is an effective model that emerges in the extended dynamical mean field theory 
 of the multipolar Kondo lattice (see the SI, Sec.\,B).
The model involves the local degrees of freedom, containing both the spin $\sigma$ and the orbital $\tau$ 
 components, which are coupled to the fermionic and bosonic baths. The former couplings describe
 the (fermionic) Kondo effect, while the latter describes the collective fluctuations associated with
 the RKKY interactions. 
 An outstanding question is whether a generic tuning trajectory leads to 
 two-stage  transitions 
 or whether it could also involve a one-stage transition. 
 We focus on 
 a multipolar Bose-Fermi Kondo model
 that arises as an effective Hamiltonian of the multipolar Kondo lattice model through 
 the extended dynamical mean field theory
 \cite{Hu-edmft2022.3,si-smith1996,smith2000spatial,chitra2001}
 (SI, Sec.\,B). In addition, we
 take advantage of the understanding on spin-only systems,
namely the emergence of new fixed points  in the Bose-Fermi Kondo model is insensitive 
 to the spin symmetry \cite{Hu-edmft2022.3,Zhu2002} and, furthermore,
  the Kondo destruction fixed points
 of the Bose-Fermi model are realized in the corresponding Kondo lattice model through 
 the extended dynamical mean field analysis.  Accordingly,
 we i) will analyze the multipolar Bose-Fermi Kondo model in its Ising-anisotropic case
 to allow for comprehensive analytical studies, 
 though we expect that the conclusion that new fixed points develop in this model will
 apply
  to the 
 spin-isotropic case as well and ii) expect that the new fixed points that we identify
 in the multipolar Bose-Fermi Kondo model will be realized as Kondo destruction quantum
 critical points in the multipolar Kondo lattice model.

We thus
 study the multipolar Bose-Fermi Kondo model at zero temperature.
By using a Coulomb-gas representation of the Bose-Fermi Kondo model,
we carry out analytical renormalization-group (RG) calculations
that are controlled by an expansion in terms of a small quantity $\epsilon$ (defined in Eq.\,\ref{bosonic-bath-spectrum}).
We uncover an overall phase diagram at zero temperature, 
which reveals the mechanism for the sequential Kondo destruction and shows that it appears for any generic 
trajectory in  the phase diagram.
Our asymptotically exact theory  points to 
a new design principle for beyond-Landau quantum criticality and strange metallicity
 in a variety of
 other strongly correlated systems,
including $d$-electron-based flat band systems and moir\'{e} structures.
\\

\begin{figure}[t]
%\centering
%\includegraphics[width=0.85\columnwidth]{fig_bare.pdf}
%\vskip 0.15in
\includegraphics[scale=1.0]
%{phasediagramRG2.png}
{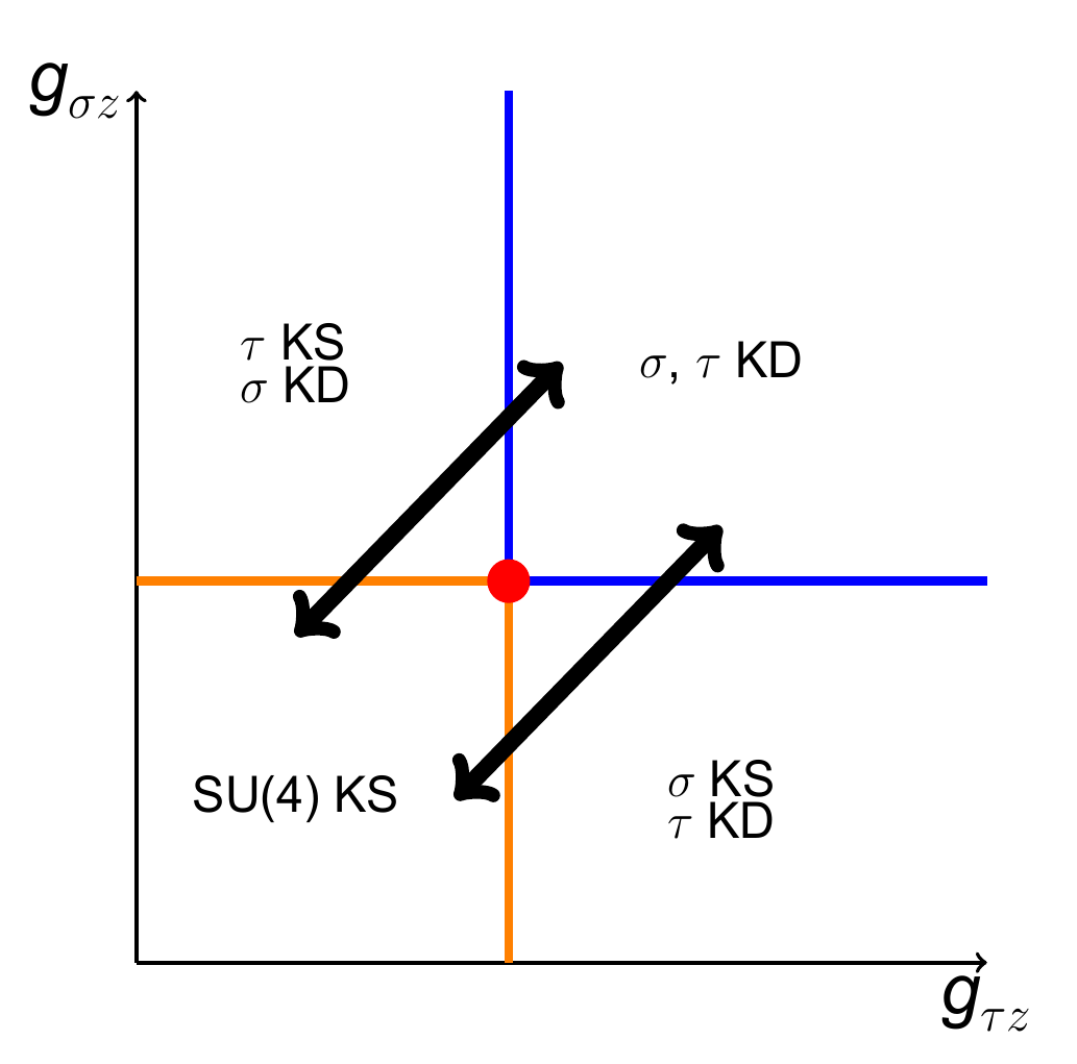}
\vskip 0.05in
\caption{
{\bf The overall phase diagram.}
Presented here is the phase diagram
in the $g_{\sigma z}$-$g_{\tau z}$ parameter 
space, for fixed Kondo couplings,
 of the model given in Eq.\,\ref{SU4RGmodel}.
KD and KS refer to the phases with Kondo-destruction and Kondo screening, 
 respectively, whereas $\sigma$ and $\tau$ refer to spin and orbital (c.f. Fig.\,\ref{fig:su4bfk2}),
 respectively.
 The black arrows mark generic trajectories in the parameter space that 
 correspond to the tuning of a non-thermal physical control parameter.
 The overall phase diagram implies two-stages of Kondo destruction along any 
 generic tuning  trajectory.}
\label{fig:su4twostage}
\end{figure}

 \noindent
{\bf Results}
\\
{\bf Sequential Destruction of Multipolar Kondo Entanglement.}
Our key findings are visualized 
 in terms of an overall phase diagram presented in
the $g_{\sigma z}$-$g_{\tau z}$ parameter space, as illustrated in Fig.\,\ref{fig:su4twostage}.
Here, $g_{\sigma z}$ and $g_{\tau z}$ are the couplings of the local multipolar moment to the bosonic fields in the 
spin and orbital channels, respectively. The fermionic Kondo couplings are kept fixed. 
Our main results are as follows:
\begin{itemize}

\item 
In the special case with the spin and orbital bosonic couplings being equal, $g_{\sigma z}=g_{\tau z}$, we identify a critical
fixed point that is accessible by the $\epsilon$-expansion. This critical point, marked by the red point in Fig.\,\ref{fig:su4twostage}, describes a one-stage transition for the destruction of the SU(4) spin-orbital Kondo effect. 

\item 
We find that the anisotropy between these two bosonic couplings is relevant in the RG sense. 
This implies that the one-stage Kondo-destruction cannot describe the quantum phase transition
along a generic trajectory in the phase diagram. 

\item 
Moreover, we are able to determine the complete phase diagram asymptotically exactly, as shown 
in Fig.\,\ref{fig:su4twostage}. This is made possible by realizing that all the phase boundaries meet at the 
equal-bosonic-coupling critical fixed point, near which the run-away RG flows are still small within the $\epsilon$-expansion.
It is further substantiated by a more comprehensive RG analysis  presented in the SI (Secs.\,C,D).
\end{itemize}

The overall phase diagram implies
two stages of Kondo-destruction QCPs for \textit{any} generic tuning trajectory at zero temperature,
 one each in the spin and orbital channels despite their entwining in the Hamiltonian.
This is illustrated by the sequence of quantum phase transitions along the solid black lines in Fig.\,\ref{fig:su4twostage}.

\noindent
{\bf Model and Solution Methods.}
We now specify the model and describe the setup for our asymptotically exact analysis.
 The multipolar Bose-Fermi Kondo model,
 schematically described by 
Fig.\,\ref{fig:su4bfk2},
  is given by the following Hamiltonian:
\begin{equation}\label{SU4RGmodel}
H_{\mathrm{BFK}}=H_{\mathrm{0}}+H_{\mathrm{K,0}}+H_{\mathrm{BK}} \   .
\end{equation}
Here, $H_{\mathrm{0}}$ is  the non-interacting part for the conduction electron $c_{p,i\alpha}$ 
and the bosonic baths $\vec{\phi}{\kappa,q}$ (where $\kappa=\sigma,\tau, m$):
\begin{equation}
\label{SU4RGmodel-H0}
\begin{aligned}
%&
H_{\mathrm{0}}=\sum_{p,i\alpha}\epsilon_{p}c^{\dagger}_{p,i\alpha}c_{p,i\alpha}
%\\
%&~~~~
\,+\,\sum_{q}W_{q}\left(\vec{\phi}_{\sigma,q}^{\dagger}\cdot\vec{\phi}_{\sigma,q}
+\vec{\phi}_{\tau,q}^{\dagger}\cdot\vec{\phi}_{\tau,q}+\vec{\phi}_{m,q}^{\dagger}\cdot\vec{\phi}_{m,q}\right)
\  .
\end{aligned}
\end{equation}

To set up controlled RG calculation, we introduce 
an expansion parameter $\epsilon$, which is defined through the bosonic spectrum $W_q$:
\begin{equation}
\label{bosonic-bath-spectrum}
\sum_{q}\left[\delta\left(\omega-W_q\right) - \delta\left(\omega+W_q\right)\right]
=\left(\frac{ K^2_0}{\pi} \right)\vert \omega \vert^{
1-\epsilon}\sgn{\omega} \ ,
\end{equation}
with $0<\epsilon<1$,
and
for $\vert \omega\vert<\Lambda$,
which specifies a high-energy cut-off scale.
The  fermionic Kondo coupling between the local multipolar moment and
 conduction electrons is as follows:
\begin{equation}\label{SU4KondoRG}
H_{\mathrm{K,0}}=\left[ J_{\sigma}\vec{ \sigma} \cdot \vec{\sigma}_{c}
+ J_{\tau} \vec{\tau} \cdot \vec{\tau}_{c} 
+ 4J_{M} \left( \vec{\sigma}_{i} \otimes \vec{\tau} \right) \cdot \left( \vec{\sigma}_{c} \otimes \vec{\tau}_{c} \right)\right]
\ ,
\end{equation}
where $\vec{\sigma}\,(\vec{\tau})$ and $\vec{\sigma}_c\,(\vec{\tau}_c)$ are the spin (orbital) operators 
of the single impurity and the conduction electrons, respectively. 
Further details and definitions are given in the Methods (Sec.\,I).

Finally, the coupling between the 
local multipolar moment and the bosonic bath is given by:
\begin{equation}\label{bkcouple2}
H_{\mathrm{BK}}=g_{\sigma z} \sigma_{z}
\phi_{\sigma z} + g_{\tau z} \tau_{z} \phi_{\tau z} +g_m\left(\sigma_z\otimes\tau_{z}\right)\phi_{m} \ ,
\end{equation}
where $\vec{\phi}_{\kappa}=\sum_{q}\left(\vec{\phi}_{\kappa,q}+\vec{\phi}^{\dagger}_{\kappa,-q}\right)$ with $\kappa=\sigma,\tau,m$.
We focus on the Ising-anisotropic case for the couplings in both the spin and orbital channels ($g_{\sigma z}$ and 
$g_{\tau z}$, respectively) as well as for the spin-orbital mixed coupling ($g_m$).
The BFK model $H_{BFK}$ (Eq.\,\ref{SU4RGmodel}) is mapped from a multipolar Kondo lattice model that contains 
a lattice of local levels with a four-fold degeneracy by the scheme of extended dynamical mean field theory
\cite{Hu-edmft2022.3,si-smith1996,smith2000spatial,chitra2001}. 
 
\begin{figure}[b!]
%[htbp]
   \centering
   \includegraphics[scale=0.28]
   %{figure3.pdf}
   {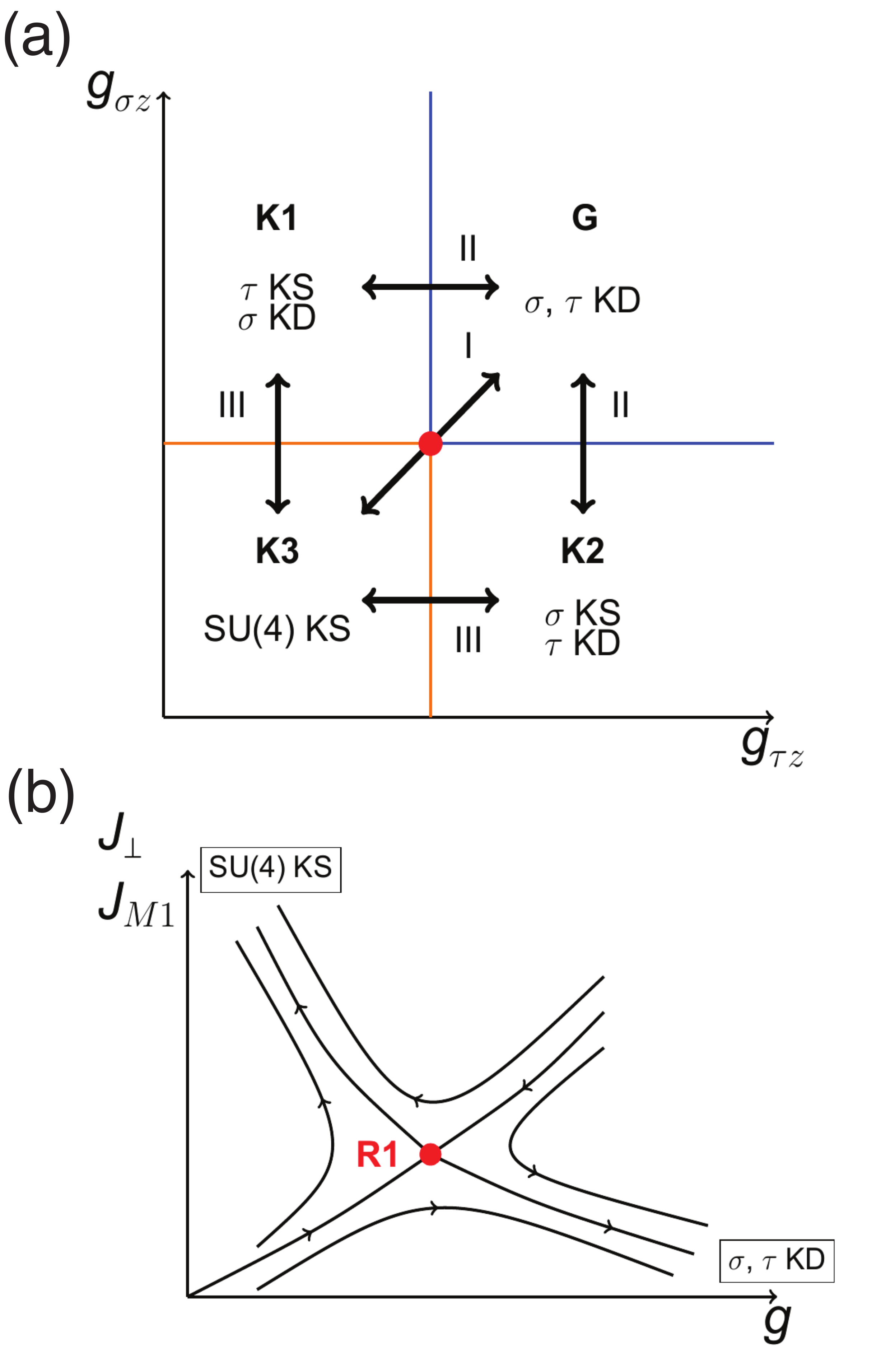}
   \caption{   
{\bf Renormalization-group analysis.}
(a) Trajectories in the parameter space of the BFK model (Eq.\,\ref{SU4RGmodel}), marked as 
``\rom{1}"-``\rom{3}", along which the RG analyses are carried out in steps.
The labels ``G", ``K1", ``K2" and ``K3" describe the RG fixed points for the corresponding phases.
(b) RG flow diagram of the reduced beta functions (Eq.\,\ref{reduceSU4}), where $g=g_{\sigma z}=g_{\tau z}$.
``\textbf{R1}" marks the unstable fixed point that captures the transition along the fine-tuned trajectory ``\rom{1}" of (a).}
\label{fig:su4RGre}
\end{figure}

We now summarize how to set up the framework
to tackle this rich problem using the (asymptotically exact)
RG approach.
We aim to determine the generic phase diagram in the $g_{\sigma z}$-$g_{\tau z}$ parameter space.
In other words, we fix the fermionic Kondo couplings and vary $g_{\sigma z}$ and  $g_{\tau z}$,
and we can keep the mixed bosonic coupling $g_m=0$ (see the Methods, Sec.\,I).
We are able to set up systematic RG calculations using a Coulomb-gas representation,
as described in some detail in the Methods (Sec.\,I).
We achieve this by dividing the analysis 
into two steps. First, we analyze the problem
along a fine-tuned trajectory in the phase diagram: along
the diagonal in the $g_{\sigma z}$-$g_{\tau z}$ space, viz. the trajectory ``I" in Fig.\,\ref{fig:su4RGre}(a).
This analysis leads to an anchoring point, which allows us 
to determine the sequence of quantum phase transitions along generic trajectories of the phase diagram.

\noindent
{\bf Quantum phase transitions: fine-tuned case.}
We now carry out
 RG calculation of the spin-orbital coupled Bose-Fermi Kondo model (Eq.\,\ref{SU4RGmodel}).
As
 outlined
in the Methods (Sec.\,I),
we will start from trajectory ``I" 
of 
Fig.\,\ref{fig:su4RGre}(a),
which corresponds to the fine-tuned case of equal bosonic couplings in the spin and orbital
channels,
$g_{\sigma z}=g_{\tau z}$.
We demonstrate a critical point 
[marked by the red solid point in 
Fig.\,\ref{fig:su4RGre}(a)]
that is accessible by 
an $\epsilon$-expansion  in our RG analysis. It describes a direct transition from
 the spin and orbital Kondo-destroyed (KD) phase  to the fully
 (spin or orbital)  Kondo-screened (KS) phase.
It will be shown in the next section that, by analyzing the vicinity of this critical point, we can 
 determine the structure of the overall phase diagram.

Generally the total number of coupling constants is seven 
(see the Methods, Sec.\,I). 
However, under the trajectory $g_{\sigma z}=g_{\tau z}=g$, 
some of the coupling constants are irrelevant, 
or can be combined due to the symmetry constraint, 
and thus the numbers of relevant RG equations (the beta functions) 
is substantially reduced. We leave the details in the SI (Sec.\,D),
and present the final reduced beta functions and their analysis in the Methods (Sec.\,II).
The RG beta functions are expressed in terms of 
 $y\propto J_{\sigma\perp}=J_{\tau\perp}$, which flips either spin or orbital indices, 
 $M\propto g^2_{\sigma z}=g^2_{\tau z}=g^2$, and $y_1\propto J_{M1}$, which is the part of the Kondo coupling  $J_M$ 
 that flips both the spin and orbital indices
(see Eq.\,\ref{kondoapp} in 
the SI, Sec.\,C).

From these reduced beta functions, Eq.\,\ref{reduceSU4},
we identify a critical point marked by the red dot in Fig.\,\ref{fig:su4RGre} and labeled as 
\textbf{R1}. Importantly, this fixed point is accessible by our $\epsilon$-expansion.
It has one relevant direction and separates the spin and orbital KD phase 
 from the SU(4) KS phase, which we call $\textbf{K3}$ for latter convenience.
Because $\textbf{R1} $ is accessible by the $\epsilon$-expansion, 
we can address what happens in the vicinity of this fixed point.
We will show in the next section that any small asymmetry 
between $g_{\tau z}$ and $g_{\sigma z}$ around \textbf{R1}  
is relevant  in the RG sense.
As a result, the direct phase transition between spin and orbital KD phase and SU(4) KS phase 
is  fine-tuned.
In other words, this direct transition occurs at a point in the parameter space  --
the red dot in Fig.\,\ref{fig:su4RGre}(a,b)
 -- instead of through a boundary line.

\noindent
{\bf Quantum phase transitions: generic cases.}
So far we have considered the case of equal bosonic couplings in the spin and orbital channels, 
\textit{i.e.}, $g_{\sigma z}=g_{\tau z}=g$. However, these two couplings are generically different.
Thus, we have to determine the quantum phase transitions along trajectories away from the diagonal
in the $g_{\sigma z}$-$g_{\tau z}$ parameter space.
We find that there are two sets of trajectories, which are marked by ``II" and ``III" in
Fig.\,\ref{fig:su4RGre}(a).
We describe our analyses of these two cases in turn.
%\vskip 0.2 cm

We next consider the transition between the spin and orbital KD phase and the spin or orbital KS phase. 
Importantly, 
we do so by starting from
the RG trajectory around the critical point \textbf{R1} 
where $g_{\sigma z}=g_{\tau z}=g^*$
between the spin and orbital KD and the SU(4) KS phases.
As we have just alluded to, around \textbf{R1},
any small asymmetry between $g_{\tau z}$ and $g_{\sigma z}$ is relevant 
with the scaling dimension $\sqrt{2\epsilon}$ (up to the order $\sqrt{\epsilon}$) in RG sense,. 

Consider first the case with a slight increase of the coupling constant $g_{\tau z}$, while keeping all the other 
parameters fixed;
in other words, now $g_{\tau z}> g_{\sigma z}=g^*$. The RG trajectory will flow towards $g_{\tau z}\rightarrow \infty$. 
We can then vary $g_{\sigma z}$ to map out the RG flow.  
The corresponding trajectory in the phase diagram 
are denoted as arrow (\rom{2}) in 
Fig.\,\ref{fig:su4RGre}(a).
Along this trajectory, the
 reduced beta functions
 are determined (see the Methods, Eq.\,\ref{reducespin}) in terms of
 $y_2 \propto J_{\sigma\perp}$, which flips only the orbital indices, and $M^{\sigma}\propto g^2_{\sigma z}$. 

From the reduced beta functions (Eq.\,\ref{reducespin}), one can identify another critical fixed point 
$\left(y^*_2,M^{\sigma *}\right)=\left(\frac{\sqrt{\epsilon}}{2},1\right)$. This fixed point 
has one relevant direction with scaling dimension $\sqrt{2\epsilon}$ (up to the order $\sqrt{\epsilon}$)
 and separates the spin KS phase $\left(y_2\rightarrow \infty,M^{\sigma }\rightarrow 0\right)$ from the spin 
 and orbital KD phase $\left(y_2\rightarrow 0,M^{\sigma }\rightarrow \infty \right)$. 

\begin{figure}[t]
%[h]
%\centering
%\includegraphics[width=1.03\columnwidth]{fig_materials.pdf}
%\vskip 0.15 in
\includegraphics[scale=0.60]
%{colombgasRG4.png}
{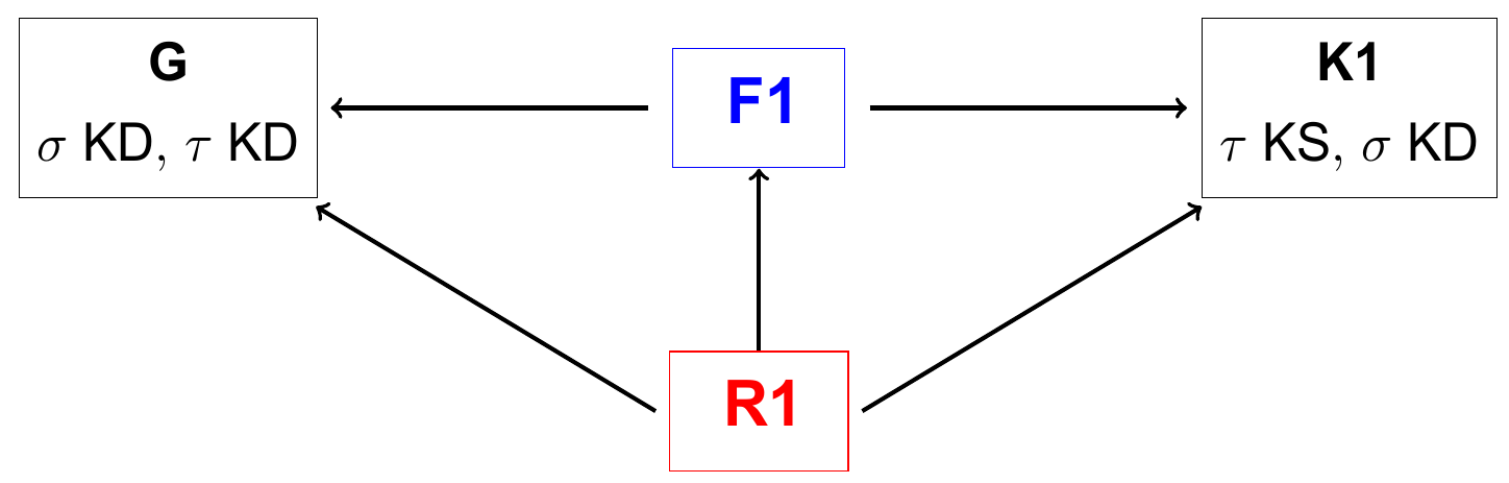}
\vskip 0.15in
\caption{{\bf The schematic renormalization-group flow structure.}
Illustrated here is the RG flow structure 
of the phase transition between the orbital and spin KD phase and 
 the spin KS phase denoted as \textbf{G} and \textbf{K1}, respectively.  
 Around the multi-critical point \textbf{R1},
    once the $g_{\sigma z}$ is slightly enlarged, the RG trajectory will flow toward \textbf{F1}, 
    which is the generic critical point separating \textbf{G} and \textbf{K1}.
}
 \label{fig:relativeRGmain}
\end{figure}

The schematic RG flow structure is shown in Fig.\,\ref{fig:relativeRGmain}, 
where the spin and orbital KD phase and the spin KS phase, denoted as \textbf{G} and \textbf{K1}, respectively,
are separated by the critical point denoted as \textbf{F1}.  
Based on this RG structure, we establish the transition between the spin and orbital KD phase and the spin KS phase. 
By applying a precisely parallel analysis, we establish the phase transition between spin and orbital KD phase 
and the orbital KS phase; we name the associated critical point as \textbf{F2}.

We have so far analyzed the transitions out of the spin and orbital KD phase.
This phase can transit into 
the spin or orbital KS phase without fine-tuning the parameters. It can also transit into the SU(4) KS phase by
fine-tuning the parameters. 

Because the spin or orbital KS phase and the SU(4) KS phase correspond to different 
stable strong coupling fixed points, there must be other generic critical points that separate them. 
These generic critical points 
describe the phase transition between the spin or orbital KS phase and the SU(4) KS phase, 
as shown in the phase diagram trajectory denoted as the dashed arrow (\rom{3}) in 
Fig.\,\ref{fig:su4RGre}(a). Here
 we would like to 
 finally establish the transition between the spin or orbital KS phase and the SU(4) KS phase, which corresponds to the trajectories (\rom{3}) in
 Fig.\,\ref{fig:su4RGre}(a).

Again, we focus on the RG trajectory around the critical point \textbf{R1} where $g_{\sigma z}= g_{\tau z}=g^*$ 
between the spin and orbital KD and SU(4) KS phases. If we keep all the other parameters fixed 
but just slightly decrease the coupling constant $g_{\sigma z}$, that is, $g_{\sigma z}< g_{\tau z}=g^*$, 
then the RG trajectory will flow towards $g_{\sigma z}\rightarrow 0$. We can then vary $g_{\tau z}$ to explore the RG trajectory.  
The corresponding trajectories in the phase diagram are denoted as the arrow \rom{3} in 
Fig.\,\ref{fig:su4RGre}(a).

However, unlike \textbf{R1} and \textbf{F1}, the real locations of the  \textbf{X1} is harder to identify directly from the 
beta functions.
To proceed, we exploit 
the property of the critical point \textbf{R1} that we alluded to earlier:
Here, all of the fugacity $y$ is $\sim \sqrt{\epsilon}$ around \textbf{R1}.
Near
 \textbf{R1}, one can thus neglect
 in a controlled way the higher order terms of $\sqrt{\epsilon}$ in the beta
  functions of the fugacity
(See 
the SI, Sec.\,D
for more details), and in the end the 
reduced beta functions 
are determined (see the Methods, Eq.\,\ref{reduce4ms})
in terms of $y_1$, $M^{\tau}$ and
$y_3\propto J_{\tau\perp}$. 

\begin{figure}[t]
%[h]
%\centering
%\includegraphics[width=1.03\columnwidth]{fig_materials.pdf}
%\vskip 0.15 in
\includegraphics[scale=0.60]
%{colombgasRGnew.png}
{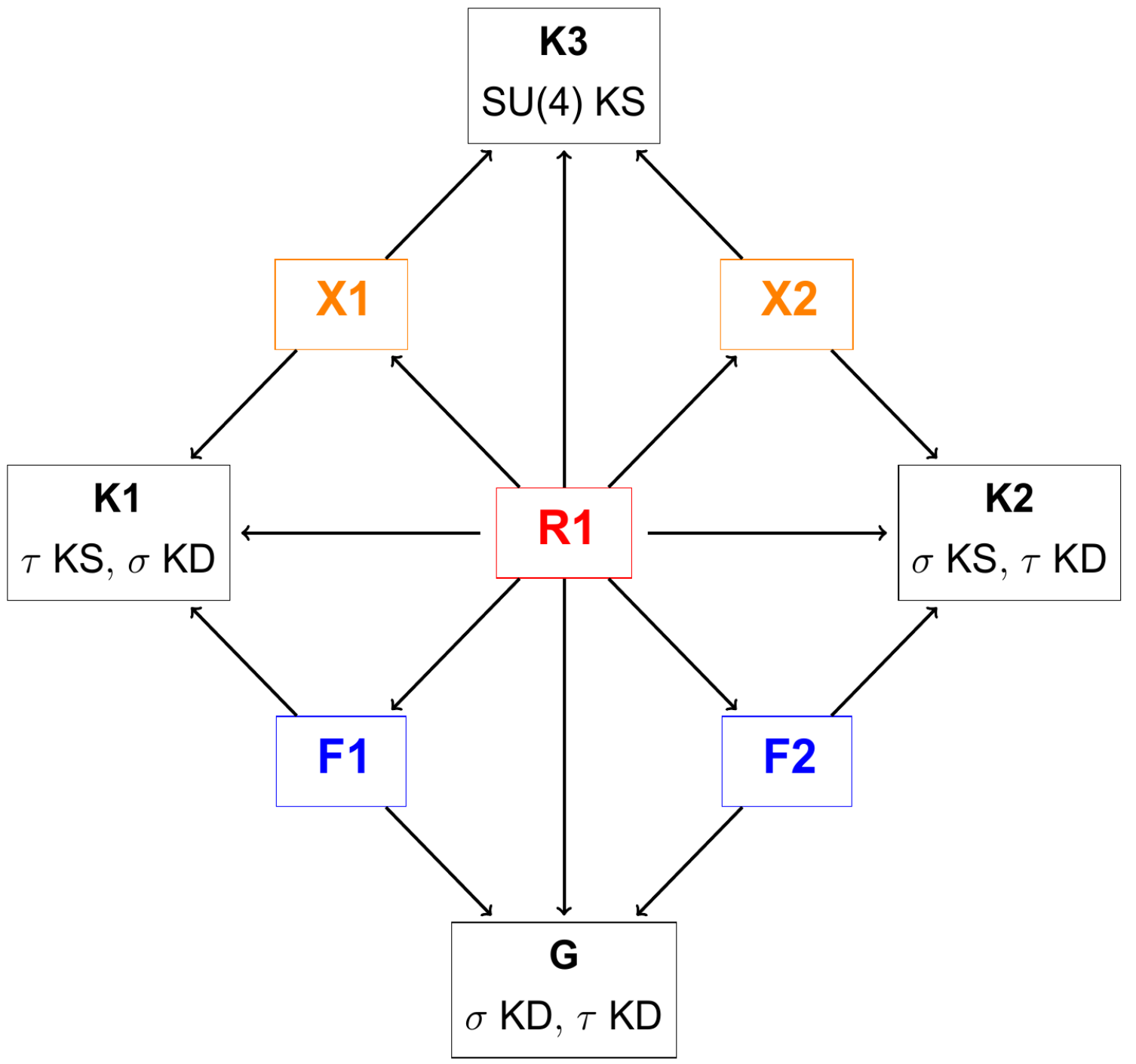}
\vskip 0.15in
\caption{
 {\bf  The schematic structure of the fixed points.}
 Illustrated here are the structure of the fixed points and the relative RG flow 
 of the BFK model (Eq.\,\ref{SU4RGmodel}), as derived 
   from the RG analysis. 
KS and KD denote the Kondo-screened and Kondo-destroyed fixed points, respectively. 
The boxes  \textbf{K1}-\textbf{K3} are different kinds of strong Kondo coupling fixed points, 
and the box \textbf{G} is the spin and orbital KD fixed point. 
The red box \textbf{R1} is a multi-critical point between spin and orbital KD phase and SU(4) KS phase.  
The blue boxed \textbf{F1}-\textbf{F2} are generic critical point separating different types the spin and orbital 
KD phases to either spin or orbital KS phases. Because the strong Kondo coupling fixed points 
\textbf{K1}, \textbf{K2}, and \textbf{K3} are stable fixed points,
they are separated by the generic critical points, denoted as orange boxes \textbf{X1} and \textbf{X2}.
}
\label{fig:su4RGresultmain}
\end{figure}

From the reduced beta functions (Eq.\,\ref{reduce4ms}),  
we  identify a critical line $\left(y^*_1, y^*_3,M^{\tau *}\right)=\left(a,\frac{\sqrt{\epsilon-4a^2}}{2},1\right)$
where $a$ is a constant,  which separates the spin and orbital KS phase from the spin KS phase 
and corresponds to the critical point \textbf{X1} in Fig.\,\ref{fig:su4RGresultmain} with scaling dimension $\sqrt{2\epsilon}$.  
By a parallel analysis, the transition between the spin and orbital KS phase and the orbital KS phase can also be established.

\noindent
{\bf Phase diagram and the sequential Kondo destruction.}
Based on the above,
we have established the 
overall phase diagram, 
which is shown in 
 Fig.\,\ref{fig:su4twostage}.
This phase diagram is also seen through a complete RG flow,
Fig.\,\ref{fig:su4RGre}(a),
 which combines the RG flows along the various trajectories we have described 
 in the previous sections. (A complementary, and more comprehensive, way of deriving this complete 
 RG flow is given in the SI, Sec.\,D.)
We summarize the characterization of the various phases and their transitions as follows:
 \begin{itemize}
 \item  
 The boxes  \textbf{K1}-\textbf{K3} are different kinds of strong Kondo coupling fixed points, 
 and the box \textbf{G} is the spin and orbital KD fixed point. 
 These fixed points are all stable according to the beta functions (Eq.\,\ref{BFKBeta}), 
 and thus describe phases of matter.
 \item
The red box \textbf{R1} is a multi-critical point between 
 the spin and orbital KD phase and SU(4) KS phase since there are two relevant directions around it.
 \item
The blue boxed \textbf{F1}-\textbf{F2} are generic critical point separating different types the spin and orbital KD phases to either spin or orbital KS phases.
 \item
Because the strong Kondo coupling fixed points \textbf{K1}, \textbf{K2}, and \textbf{K1} are stable fixed points,
they are separated by the generic critical points \textbf{X1} and \textbf{X2}.  
The generic critical points \textbf{X1} and \textbf{X2} control the critical phenomena of the trajectories \rom{3} 
in 
Fig.\,\ref{fig:su4RGre}(a).
 \end{itemize}
 
The solid black arrows in  Fig.\,\ref{fig:su4twostage} marks the generic tuning trajectories in the zero-temperature
 phase diagram. Along each of such trajectories, two-stages of Kondo destruction take place, each characterizing a QCP
 in the spin or orbital channel. This asymptotically exact result
 provides a firm theoretical basis to understand the 
 field-induced quantum phase transitions that have been experimentally observed
  in 
   Ce$_3$Pd$_{20}$Si$_6$ \cite{Martelli2019}.

\noindent
{\bf Discussion}
\\
In this work, we have performed a detailed renormalization-group
 analysis of a spin-orbital-entwined Bose-Fermi-Kondo model,
which is mapped from a multipolar Kondo lattice model. 
We are able to determine the overall phase diagram at zero temperature,
which reveals the mechanism for the sequential Kondo destruction and shows that
it appears for any generic 
trajectory in  the phase diagram. As such, our results provide a firm theoretical basis for 
understanding the surprising experimental results 
in the heavy fermion metal Ce$_3$Pd$_{20}$Si$_6$  \cite{Martelli2019}.
More generally, our work elucidates the quantum criticality in spin-orbital-coupled heavy fermion systems.

Our asymptotically exact theoretical results also make it clear how 
the entwining of spins, orbitals and other quantum numbers in local degrees of freedom
allows for new types of
quantum criticality and 
associated strange metallicity.
This represents a design principle for creating and realizing new forms of quantum criticality and associated 
strange metallicity:
The cooperation of strong correlations,
large spin-orbit coupling and crystalline symmetry represents a robust means to create varied local degrees of freedom;
and the tuning of such strongly correlated systems can realize a sequence of beyond-Landau quantum critical points.
Beyond heavy fermion metals, effective local degrees of freedom have also been advanced 
for pertinent
molecular orbitals of 
$d$-electron-based flat band systems 
\cite{Ye2021.x,Ekahana2022.x,HaoyuHu2022.x,LeiChen2022.x}
 and moir\'{e} states of twisted structures \cite{Ramires2021,Song2022,Guerci2022.x}.
Thus,
we expect this design procedure to operate 
not only in multipolar heavy fermion metals, but also in 
transition-metal 
compounds,
synthetic systems such as 
moir\'{e} structures 
and beyond.

{\it Note added.} The sequential Kondo destruction that we identify in the minimal prototype 
multipolar model has now 
also been seen in a related model that contains additional couplings
and has  continuous spin symmetry
 (S. E. Han, D. J. Schultz and Y. B. Kim, 
``Microscopic theory of multi-stage Fermi surface reconstruction in higher-rank moment quantum materials").

\vskip 1 cm

%%%%%%%%%%%%%%
\noindent{\bf\large Methods}\\
%\\
\noindent
{\bf I.~Model and the Renormalization-Group Method.}
In the definition of the model, 
Eqs.\,\ref{SU4RGmodel}-\ref{SU4KondoRG},
the spin and orbital operators of the conduction electrons are defined as:
 \begin{equation}
\begin{aligned}
&\vec{\sigma}_c 
= \frac{1}{2} \sum_{i,\alpha\beta} c^{\dagger}_{i\alpha} {\vec{ s}}_{\alpha \beta} 
c_{i\beta  } \ ,
\\
&\vec{\tau}_c
= \frac{1}{2} \sum_{ij, \alpha } c^{\dagger}_{i \alpha } {\vec{t }}_{i j } 
c_{j\alpha } \ ,
\\
&\vec{\sigma}\otimes \vec{\tau}_c
= \frac{1}{4} \sum_{ij,\alpha\beta } c^{\dagger}_{i\alpha } {\vec{s}}_{\alpha\beta} \otimes{\vec{t }}_{ij } 
c_{j\beta } \ .
\end{aligned}
\end{equation}
Here, we use  $\alpha, \beta$ and $i,j$
 to denote the spin and orbital indices, respectively.
 Thus, ${\vec{ s}}_{\alpha\beta}$ and ${\vec{ t }}_{ij}$ 
are Pauli matrices in the spin and orbital subspaces, respectively.
For the fermionic Kondo Hamiltonian alone, the anisotropy in the 
couplings is generically unimportant as the system restores
the SU(4) symmetry in the Kondo-entangled ground state \cite{Hewson-book}.
We have therefore chosen the bare
Kondo Hamiltonian to be SU(2) symmetric in the spin as well as in the orbital sector, with an 
overall SU(2)$\otimes$SU(2) symmetry.  
The full renormalized Kondo Hamiltonian (Eq.\,\ref{kondoapp}) for the later RG analysis 
is shown in the SI, Sec.\,C.
%Appendix (\ref{appen:RGCoulomb}).

We now describe the framework
to tackle this rich problem using the (asymptotically exact)
RG approach.
 Further details can be found 
in the 
the SI (Sec.\,C).
 
First, 
our goal is to study the generic phase diagram in the $g_{\sigma z}$-$g_{\tau z}$ parameter space.
In other words, we fix the fermionic Kondo couplings and vary $g_{\sigma z}$ and  $g_{\tau z}$. 
For this purpose, it suffices to keep the mixed bosonic coupling $g_m=0$.
A non-vanishing but small $g_m$  does not modify the structure of the
phase diagram, as we show in 
the SI (Sec.\,E).
To proceed, we use a 
bosonization approach to represent the BFK model (Eq.\,\ref{SU4RGmodel}) 
in terms of a Coulomb gas, from which a controlled RG calculation 
based on an expansion in $\epsilon$
is possible\cite{Zhu2002,Qimiao1996, Smith1999}. 
We note that the Coulomb-gas RG calculation is based on a dilute-instanton expansion,
which is non-perturbative 
in stiffness constants but perturbative in terms of fugacities \cite{Zhu2002}.

Second, the Ising couplings of $H_{BK}$ (Eq.\,\ref{bkcouple2}) break not only the SU(4) symmetry 
but also the smaller SU(2)$\times$SU(2) symmetry. 
While the Kondo couplings in $H_{K}$ respect the SU(2)$\times$SU(2) symmetry, under the RG flow these couplings 
will generically become spin anisotropic.
It turns out that one needs to consider five types of Kondo couplings. Together with the spin and orbital Ising couplings 
$g_{\sigma z}$ and $g_{\tau z}$ of $H_{BK}$ (Eq.\,\ref{bkcouple2}), the total number of RG coupling constants is seven. 
The large number of the RG charges makes it a challenge to  determine the overall RG flow structure. 
We are able to accomplish this goal by analyzing the problem in several steps.

Crucially, we take the first step to be  a fine-tuned trajectory in the phase diagram.
Recall that the $g_{\sigma z}$-$g_{\tau z}$ parameter space is of our interest.
For clarity, we visualize this parameter space in 
Fig.\,\ref{fig:su4RGre}(a),
which 
marks the relevant phases. The fine-tuned trajectory  we focus our initial analysis on
corresponds to identical couplings to the bosonic baths in the spin and orbital sectors.
It goes along the diagonal in the $g_{\sigma z}$-$g_{\tau z}$ space, and is marked as trajectory ``I".
The result of the analysis on this fine-tuned trajectory provides a anchoring point, which allows us 
to determine the sequence of quantum phase transitions along generic trajectories of the phase diagram.

We note that it is possible to rigorously establish the phase diagram, 
Fig.\,\ref{fig:su4RGre}(a),
through a comprehensive RG analysis without taking the 
fine-tuned trajectory ``I" as the starting anchoring
point. This is described in
the SI (Secs.\,C,D).
We choose to present the step-by-step analysis here in the main text,
given that it reveals the underlying physics in a considerably more transparent way.

\noindent
{\bf II.~RG equations and analysis: fine-tuned case.}
The RG analysis, described in the SI (Sec.\,D), leads to 
the following reduced beta functions:

\begin{equation}\label{reduceSU4}
\begin{aligned}
&\frac{dy_{1}}{dl}=\left(1-2M\right)y_{1}+2y^2 \ , \\
&\frac{dy}{dl}=\left(1-M\right)y+2y_1y \ , \\
&\frac{dM}{dl}=\left(\epsilon -4y^2_1-4y^2\right)M \  .\\    
\end{aligned}
\end{equation}
Note that we can set $J_{\sigma\perp}=J_{\tau\perp}$,
given that we are considering
a path in the parameter space that preserves the symmetry $\sigma\leftrightarrow \tau$. 
From these reduced beta functions (Eq.\,\ref{reduceSU4}),
we identify a critical point (up to the order $\sqrt{\epsilon}$) at
$\left(y^*_1,y^*,M^*\right)=\left(\frac{-1+\sqrt{1+12\epsilon}}{12},\frac{\sqrt{-1+12\epsilon
+\sqrt{1+12\epsilon}}}{6\sqrt{2}},\frac{5+\sqrt{1+12\epsilon}}{6}\right)\cong\left(0,\frac{\sqrt{\epsilon}}{2},1\right)$.
This fixed point has one relevant direction with the scaling dimension $\sqrt{2\epsilon}$
and separates the spin and orbital KD phase $\left(y_1\rightarrow 0,y\rightarrow 0,M\rightarrow \infty\right)$ 
 from the SU(4) KS phase $\left(y_1\rightarrow \infty,y\rightarrow\infty,M\rightarrow 0\right)$\cite{ignore}. 
 The RG flow structure of the reduced beta functions (Eq.\,\ref{reduceSU4}) is plotted in Fig.\,\ref{fig:su4RGre}.  
 For latter convenience, we name SU(4) KS phase as $\textbf{K3}$, and the critical point (the red dot in Fig.\,\ref{fig:su4RGre})
 as \textbf{R1}. 

For our analysis, one feature of the fixed point $\textbf{R1} $ plays a crucial role. 
While the fixed-point value for the RG charge 
$M$ is O($1$), the values for the RG charges (the fugacities) $y_1$ and $y$ are of order $\sqrt{\epsilon}$. 
Because of this feature, the quardratic-in-$y_{\alpha}$ terms in the beta-functions of the fugacities
 turn out to be unimportant for both RG flow structure and the leading order of the scaling dimensions.
 The same conclusion is also seen in the scaling dimensions of the RG variables near 
 $\textbf{R1} $; to the leading non-vanishing order in $\epsilon$, they are the same regardless of whether 
 the quardratic-in-$y_{\alpha}$ terms are kept in the beta-functions of the fugacities.
In the next section, we'll see how this allows us to determine the 
overall structure of the phase diagram by expanding the 
RG equations around the fixed point $\textbf{R1} $. 
In particular, it allows us to carry out a complete analysis of the quantum phase transition
along trajectory ``III", which otherwise would have been much harder to achieve.

\noindent
{\bf III.~RG equations for the generic case -- trajectory \rom{2}.}
Along this trajectory, the
 reduced beta functions are calculated to be as follows:
\begin{equation}\label{reducespin}
\begin{aligned}
&\frac{dy_{2}}{dl}=\left(1-M^{\sigma}\right)y_{2} \ ,\\ 
&\frac{dM^{\sigma}}{dl}=\left(\epsilon -4y^2_2\right)M^{\sigma} \ . \\  
\end{aligned}
\end{equation}
Again, we leave the details of
 the derivation to
 the SI (Sec.\,D).
 
 \noindent
 {\bf IV.~RG equations for the generic case -- trajectory \rom{3}.}
 Along this trajectory, the
reduced beta functions 
are as follows:
\begin{equation}\label{reduce4ms}
\begin{aligned}
&\frac{dy_{1}}{dl}=\left(1-M^{\tau}\right)y_{1} \ ,\\
&\frac{dy_{3}}{dl}=\left(1-M^{\tau}\right)y_{3} \ ,\\
&\frac{dM^{\tau}}{dl}=\left(\epsilon -4y^2_1-4y^2_3\right)M^{\tau} \ .\\
\end{aligned}
\end{equation}

\clearpage

%%%%%%%%%%%%%%%%%%%%
%\vspace{0.5cm}
\vspace{0.3cm}
\noindent{\bf Acknowledgments}\\
We thank Ang Cai, Kevin Ingersent, Emilian Nica  and Rong Yu
for useful discussions. The work has been supported in part by the 
NSF Grant No. DMR-2220603
and the Robert A. Welch Foundation Grant No. C-1411. 
Work in Vienna has been supported by the 
Austrian Science Fund (P29296 and 29279) 
and the European Community (H2020 Project No.  824109).
 One of us (Q.S.) acknowledges the  hospitality  of  the  Aspen  Center  for  Physics,  
 which  is  supported  by  the  NSF  grant No. PHY-1607611.

\vspace{0.2cm}
\noindent{\bf Author contributions}
\\
C.-C. L., S. P. and Q.S. conceived the research. C.-C. Liu and Q.S. carried out theoretical model studies.
S. P. and Q. S. provided insights into multipolar heavy fermion systems.
C.-C. L. and Q. S. wrote the manuscript, with input from S.P..
%\\

\vspace{0.2cm}
\noindent{\bf Competing 
 interests}\\
The authors declare no competing 
 interests.
 %\\
 
 \vspace{0.2cm}
 \noindent{\bf Additional information}\\
Correspondence and requests for materials should be addressed to 
Q.S. (qmsi@rice.edu)
% \\

 \vspace{0.2cm}

 \noindent
{\bf Data availability}\\
%~~
All data needed to evaluate the conclusions in the paper 
are presented in the paper and/or the Supplementary Information.

%\newpage

\clearpage
%%%%%%%%%%%%%
%%%%%%%%%%%%%%%%%%%%%%%%%%%%%%%%%%%%%%%
\clearpage

%%%%%%%%%%%%%
%%%%%%%%%%%%%%%%%%%%%%%%%%%%%%%%%%%%%%%
\clearpage

%%%%%%%%%%%%%%%%%%%%%%%%%%%%%%%%%%%%

\clearpage

%%%%%%%%%%%%%
%%%%%%%%%%%%%%%%%%%%%%%%%%%%%%%%%%%%

\clearpage

%%%%%%%%%%%%%
%%%%%%%%%%%%%%%%%%%%%%%%%%%%%%%%%%%%

\clearpage

\setcounter{figure}{0}
\setcounter{equation}{0}
\makeatletter
\renewcommand{\thefigure}{S\@arabic\c@figure}
\renewcommand{\theequation}{S\arabic{equation}}

\noindent{\bf\Large Supplementary Information}\\
%\\
\section*{}
\iffalse

\newpage
\begin{widetext}

\appendix
\fi

\noindent
\subsection*{A.~The case of Ce$_3$Pd$_{20}$Si$_6$}
%\label{appen:pnasresult}
In Ce$_3$Pd$_{20}$Si$_6$, every Ce$^{3+}$ ion contributes one localized 4f electron. Because of the strong spin-orbit coupling, the spin and orbital degree of freedom of the 4f electron are coupled to together into a total angular momentum $J=5/2$ state that has six-fold degeneracy and hence supports not only dipole moment but also higher-order multipolar moments. Such six-fold degeneracy is split into a $\Gamma_8$ quartet and a $\Gamma_7$ doublet due to the crystal field effect\cite{Shi1997}. The analysis of temperature dependent inelastic neutron scattering and entropy data also revealed that the $\Gamma_8$ quartet is the true ground state for the local levels\cite{Lor2012}, which can be represented in the pseudo-spin $\vec{\sigma}$ and pseudo-orbital $\vec{\tau}$ notation as:
\begin{equation}
\begin{aligned}
&|\tau_{z}=1;\sigma_{z}=1\rangle =
\sqrt{\frac{5}{6}} |J_{z}=\frac{5}{2} \rangle 
+\sqrt{\frac{1}{6}} |J_{z}=-\frac{3}{2} \rangle\ ,  \\
&|\tau_{z}=1;\sigma_{z}=-1\rangle =
\sqrt{\frac{1}{6}} |J_{z}=\frac{5}{2} \rangle \ ,
+\sqrt{\frac{5}{6}} |J_{z}=-\frac{3}{2} \rangle \ , \\
&|\tau_{z}=-1;\sigma_{z}=1\rangle = |J_{z}= \frac{1}{2} \rangle\ , \\
&|\tau_{z}=-1;\sigma_{z}=-1\rangle = |J_{z}= -\frac{1}{2} \rangle \ .
\end{aligned}
\end{equation}

The $\Gamma_8$ systems comprise dipoles, quadrupoles, and octupoles, all of which are irreducible representations 
of the $O_{h}$ group of the cubic lattice. Both dipolar (magnetic) and quadrupolar order (and likely even octupolar order) 
may arise via the RKKY interaction between the local multipolar
moments\cite{Shi1997}. 

Applying a magnetic field leads to a sequence of two QCPs, which are associated with the magnetic and quadrupolar 
degrees of freedom respectively \cite{Martelli2019}. Across each QCP,
a jump of the Hall coefficient is found based on extrapolation of its isothermal 
dependence to the zero-temperature limit\cite{Martelli2019}. Each jump implicates a destruction of Kondo effect
and the concomitant electronic localization-delocalization phase transition at zero temperature. 

\noindent
\subsection*{B.~Multipolar Kondo lattice model}
%{\label{appen:Kondolattice}}

We consider a multipolar Kondo lattice model that contains a lattice of local levels 
with a four-fold degeneracy which can be expressed in term of spin $\vec{\sigma}$ 
and orbital $\vec{\tau}$ operators:
\begin{equation}\label{multikondo}
H_{KL}= H_{c}+H_{f,I} + H_{K} \ .
\end{equation}

The first part $H_{c}=\sum_{
{\vec{ k}} \sigma \tau} \epsilon_{ {\vec{ k}} \sigma \tau} c^{\dagger}_{ {\vec{k}}
\sigma \tau} c_{ {\vec{ k}} \sigma \tau}$ defines the kinetic energy of the
conduction electrons, and the second part 
%$H_l$ 
$H_{f,I}$
describes the RKKY interaction among the $\Gamma_{8}$ local levels. 
For the purpose of convenience and demonstration, we choose 
%$H_l$ 
$H_{f,I}$
as the Ising type:

\begin{equation}\label{SU4RKKY}
%H_l
H_{f,I}=\sum_{ij} \left [ I_{ij}^{\sigma} \sigma_{i}^{z}
\sigma_{j}^{z} + I_{ij}^{\tau} \tau_{i}^{z} \tau_{j}^{z} + I_{ij}^{m}
\left(\sigma_{i}^{z} \otimes \tau_{i}^{z} \right) \left( \sigma_{j}^{z} \otimes
\tau_{j}^{z} \right) \right ]\ ,
\end{equation}
where, $\vec{\sigma}$, $\vec{\tau}$, and $\vec{\sigma} \otimes \vec{\tau}$ express the spin and orbital operators and their tensor product, respectively, and $I_{ij}^{\sigma}, I_{ij}^{\tau}, I_{ij}^{m}$ are the corresponding coupling constant. Note that here the English letter $i,j$ are indices for sites. The Hamiltonian is essentially the Ising anisotropic version of the Kugel-Khomskii model.

The final part $H_K$ is the Kondo coupling between the local levels and their conduction-electron counterparts:
\begin{equation}\label{SU4Kondo}
\begin{aligned}
&H_{K}=\sum_{i}\left[ J_{\sigma}\vec{ \sigma}_{i} \cdot \vec{\sigma}_{i,c}
+ J_{\tau} \vec{\tau}_{i} \cdot \vec{\tau}_{i,c} 
+ 4J_{M} \left( \vec{\sigma}_{i} \otimes \vec{\tau}_{i} \right) \cdot \left( \vec{\sigma}_{i,c} \otimes \vec{\tau}_{i,c} \right)\right]\ ,
\end{aligned}
\end{equation}
where the antiferromagnetic Kondo coupling $J_{\kappa}>0$ with $\kappa=\sigma,
\tau, M$, respectively, describe the interaction of the local levels $\vec{\sigma}$, $\vec{\tau}$,
and $\vec{\sigma} \otimes \vec{\tau}$ with the conduction-electron counterparts. 

The multipolar Bose-Fermi Kondo model $H_{BFK}$ (Eq.\,\ref{SU4RGmodel}) is mapped from 
the multipolar Kondo lattice model $H_{KL}$ (Eq.\,\ref{multikondo}) 
under the extended dynamical mean field theory \cite{Hu-edmft2022.3,si-smith1996,smith2000spatial,chitra2001}. 
In this procedure, all the sites except for a local impurity are
 traced out, and the effect of the RKKY interactions between the local multipolar moments 
is to act effectively as a self-consistent bosonic bath that, along with the self-consistent fermionic bath,
are coupled to the local impurity. 
Specifically, in the EDMFT procedure, the multipolar Kondo lattice is mapped to 
an effective action that contains the local Kondo couplings as well as the following retarded interactions
(where $\beta = 1/k_B T$):
\begin{eqnarray}
\Delta S_{{\rm loc},\sigma^z}
%&=&
\,=\,
% {\cal S}_{\text{top}}
%+\int_0^{\beta} d \tau ~J_K ~{\bf S} \cdot {\bf s}_{c} \nonumber\\
%&&- \int_{0}^{\beta} d \tau \int_{0}^{\beta} d \tau' \left[ \sum_{\sigma} c_{\sigma} ^ {\dagger} (\tau)G_0^{-1}(\tau - \tau ') c_{\sigma}(\tau') \right. \nonumber\\&& \left. + \; 
-{1 \over 2}
 \int_{0}^{\beta} d \tau
\int_{0}^{\beta} d \tau'
\sigma^z (\tau) \cdot \chi_{0,\sigma^z}^{-1}(\tau - \tau') \sigma^z(\tau') \, ,
%\rule[-2ex]{0ex}{4ex}\right] 
%+ \int_{0}^{\beta} d \tau h_{\text{loc}} \, S^z \, .
\label{S-imp-kondo-lattice}
\end{eqnarray}
with a self-consistency equation stating that the auto-correlation function of 
$\sigma^z$ in the effective action is the same as its local correlation
function of the Kondo lattice,
and the corresponding parts for the entwined degrees of freedom 
$\tau^z$ and 
$\left(\sigma_{i}^{z} \otimes \tau_{i}^{z} \right)$.
This action can, via a Caldeira-Leggett
procedure \cite{Caldeira-Leggett}, 
be equivalently expressed in a Hamiltonian form, leading to Eq.\,5 of the main text.
As we described in the main text, prior studies of spin-only Kondo lattice models
within the extended dynamical mean field theory have shown
 that  the Kondo destruction fixed points
 of the Bose-Fermi model without self-consistency but with subohmic ($\epsilon >0$,
 {\it c.f.} Eq.\,3 of the main text)
 are realized in the corresponding Kondo lattice model through 
 the extended dynamical mean field analysis
  \cite{Hu-qcm2022.2,Hu-edmft2022.3}.
 That the bosonic spectrum is subohmic captures the qualitative physics that
 the collective fluctuations of the Kondo lattice
makes the density of states associated with the bosonic spectrum to be enhanced compared 
with the ohmic bath induced by the free electron-hole excitations
 \cite{Hu-qcm2022.2,Hu-edmft2022.3}.

\noindent
\subsection*{C.~Derivation of the Coulomb gas action and RG equations}
%\label{appen:RGCoulomb}

We now describe the Coulomb gas action and the procedure of the RG analysis.
The Coulomb-gas action is canonical as established for the Kosterlitz-Thouless transition
for the case of a single fugacity, and is valid when the fugacity is small (though it is non-perturbative 
in 
stiffness constant).
Our case is more
complex 
because it involves several fugacities that are coupled together. 
Nonetheless, the Coulomb-gas action can still be constructed, as graphically illustrated below (Fig.\,S1),
which allows for the systematic construction of the RG equations.

As we mentioned in the main text, because of the Ising-type couplings
\begin{equation}\label{bosonapp}
H_{BK}=g_{\sigma z} \sigma_{z}
\phi_{\sigma z} + g_{\tau z} \tau_{z} \phi_{\tau z}\ ,
\end{equation}
the whole Bose-Fermi Kondo model (Eq.\,\ref{SU4RGmodel}) breaks not only the SU(4) symmetry but also the SU(2)$\times$SU(2) symmetry. Therefore, to perform the RG calculation, one need to reduce the symmetry in the Kondo part $H_{K}$ and introduce much more Kondo couplings. To our purpose, the model with the minimal number of parameters that we need to consider is:
\begin{equation}\label{apSU4RGmodel}
\begin{aligned}
&H_{BFK}=H_0+H_{K,o}+H_{BK}\ ,\\
\end{aligned}
\end{equation}
where $H_0$ is the non-interacting part for the conduction electron $c_{p,i\alpha}$ and the bosonic bath $\vec{\phi}_{\kappa,q}$($\kappa=\sigma,\tau,m$), and the Kondo coupling $H_{K,o}$ is
\begin{equation}\label{kondoapp}
\begin{aligned}
&H_{K,o}=J_{\sigma z}\sigma^z\sigma^z_c+J_{\sigma \perp}\left(\sigma^x\sigma^x_c+\sigma^y\sigma^y_c\right)+J_{\tau z}\tau^z\tau^z_c+J_{\tau \perp}\left(\tau^x\tau^x_c+\tau^y\tau^y_c\right)\\
&+4J_{M1}\left[\left(\sigma_x\otimes\tau_x\right)\left(\sigma_x\otimes\tau_x\right)_c+\left(\sigma_x\otimes\tau_y\right)\left(\sigma_x\otimes\tau_y\right)_c+\left(\sigma_y\otimes\tau_x\right)\left(\sigma_y\otimes\tau_x\right)_c+\left(\sigma_y\otimes\tau_y\right)\left(\sigma_y\otimes\tau_y\right)_c\right]\\
&+4J_{M2}\left[\left(\sigma_z\otimes\tau_x\right)\left(\sigma_z\otimes\tau_x\right)_c+\left(\sigma_z\otimes\tau_y\right)\left(\sigma_z\otimes\tau_y\right)_c\right]+4J_{M3}\left[\left(\sigma_x\otimes\tau_z\right)\left(\sigma_x\otimes\tau_z\right)_c+\left(\sigma_y\otimes\tau_z\right)\left(\sigma_y\otimes\tau_z\right)_c\right]\\
&+4J_{M4}\left[\left(\sigma_z\otimes\tau_z\right)\left(\sigma_z\otimes\tau_z\right)_c\right]\\
\end{aligned}
\end{equation}
with $J_{\sigma\perp}=J_{M3}$ and $J_{\tau\perp}=J_{M2}$. 
Note that compared with the coupling with bosnic bath (Eq.\,\ref{bkcouple2}),  we had already set $g_m=0$ in the coupling (Eq.\,\ref{bosonapp}). In 
the SI (Sec.\,E), we
will also show that a non-vanishing but small $g_m$ does not modify the structure of our phased diagram based on the RG analysis.

Note that, without the bosonic coupling (Eq.\,\ref{bkcouple2}), both the Hamiltonian (Eq.\,\ref{SU4RGmodel}) 
and (Eq.\,\ref{apSU4RGmodel}) 
admit only a SU(4) Kondo-screened  fixed point. In addition, as we will see, 
tuning the bosonic coupling (Eq.\,\ref{bosonapp})
breaks the SU(4) Kondo-screened fixed point directly down to either a spin or orbital SU(2) Kondo-screened fixed point. 
Therefore, one should expect that how the bare Kondo couplings deviate from the SU(4) symmetric case 
does not really matter, 
and the RG analysis of the model (Eq.\,\ref{apSU4RGmodel})
captures the generic phase diagram of model (Eq.\,\ref{SU4RGmodel}).   

For the Ising-type bosonic coupling, to perform a controllable RG calculation, one need to map the Bose-Fermi Kondo model into a Coulomb gas type model\cite{Zhu2002,Qimiao1996}. The first step  to decompose the above Hamiltonian $H_{BFK}$ into the part $H_0$ that is diagonal in the space of the single impurity states $\vert \sigma\rangle\otimes \vert\tau\rangle$, and the other part $H_f$ that is not:
\begin{equation}
H_{BFK}=H_D+H_f\ ,
\end{equation}
where $H_D$ is diagonal in the space of the single impurity states $\vert \sigma\rangle\otimes \vert\tau\rangle$. We use the notation $\vert m\rangle=\vert i\alpha\rangle$ to denote a single impurity state with the orbital $i=1,2$ and the spin $\alpha=\uparrow,\downarrow$. Therefore,

\begin{equation}
H_D=\sum_{m}H_{m}\vert m\rangle\langle m \vert\ .
\end{equation}

Then we rewrite $H_m$ in term of the projection operators $X_{m m}=\vert m\rangle\langle m\vert=\vert i\alpha\rangle\langle i\alpha\vert$, so that:
\begin{equation}
\begin{aligned}
&H_{m}=E_{m}+\sum_{n}V^{n}_{m}c^{\dagger}_{n}c_{n}+\sum_{k,n}E_{k}c^{\dagger}_{k,n}c_{k,n}+\sum_{q}W_{q}\left(\vec{\phi}_{\sigma,q}^{\dagger}\cdot\vec{\phi}_{\sigma,q}+\vec{\phi}_{\tau,q}^{\dagger}\cdot\vec{\phi}_{\tau,q}\right)\\
&+\sum_{q}F^{
m}_{\sigma}\left(\phi_{\sigma z,q}+\phi^{\dagger}_{\sigma z,-q}\right)+\sum_{q}F^{
m}_{\tau}\left(\phi_{\tau z,q}+\phi^{\dagger}_{\tau z,-q}\right)\ ,
\end{aligned}
\end{equation}
where
\begin{equation}
\begin{aligned}
&V^{i\alpha}_{i\alpha}=\frac{1}{4}\left(J_{\sigma z}+J_{\tau z}+J_{M4}\right)\ ,\\
&V^{i\overline{\alpha}}_{i\alpha}=\frac{1}{4}\left(J_{\tau z}-J_{\sigma z}-J_{M4}\right)\ ,\\
&V^{\overline{i}\alpha}_{i\alpha}=\frac{1}{4}\left(J_{\sigma z}-J_{\tau z}-J_{M4}\right)\ ,\\
&V^{\overline{i}\overline{\alpha}}_{i\alpha}=-\frac{1}{4}\left(J_{\sigma z}+J_{\tau z}-J_{M4}\right)\ ,\\
\end{aligned}
\end{equation}
and
\begin{equation}
\begin{aligned}
&F^{i\uparrow}_{\sigma}=g_{\sigma z}\ ,\\
&F^{i\downarrow}_{\sigma}=-g_{\sigma z}\ ,\\
&F^{1\alpha}_{\tau}=g_{\tau z}\ ,\\
&F^{2\alpha}_{\tau}=-g_{\tau z}\ .
\end{aligned}
\end{equation}
Here we use the over-line symbol to denote the complement of the spin or orbital index.

On the other hand, the flipping part is defined as:
\begin{equation}
H_{f}=\sum_{m\neq n}Q\left(m,n\right),
\end{equation}
where
\begin{equation}
Q\left(m,m\right)=\vert m\rangle\langle m\vert H_f\vert n\rangle\langle n\vert
\end{equation}
describing the process of flipping from the single impurity state $\vert n\rangle$ to $\vert m\rangle$\ . Specifically, 
\begin{equation}
\begin{aligned}
&Q\left(i\alpha,\overline{i}\overline{\alpha}\right)=J_{M1}c^{\dagger}_{\overline{i}\overline{\alpha}}c_{i\alpha}\vert i\alpha\rangle\langle \overline{i}\overline{\alpha}\vert\ ,\\
&Q\left(i\alpha,i\overline{\alpha}\right)=\frac{1}{2}\left(J_{\sigma\perp}-J_{M3}\right)c^{\dagger}_{\overline{i}\overline{\alpha}}c_{\overline{i}\alpha}\vert i\alpha\rangle\langle i\overline{\alpha}\vert+\frac{1}{2}\left(J_{\sigma\perp}+J_{M3}\right)c^{\dagger}_{i\overline{\alpha}}c_{i\alpha}\vert i\alpha\rangle\langle i\overline{\alpha}\vert=J_{\sigma\perp}c^{\dagger}_{i\overline{\alpha}}c_{i\alpha}\vert i\alpha\rangle\langle i\overline{\alpha}\vert\ ,\\
&Q\left(i\alpha,\overline{i}\alpha\right)=\frac{1}{2}\left(J_{\tau\perp}-J_{M2}\right)c^{\dagger}_{\overline{i}\overline{\alpha}}c_{i\overline{\alpha}}\vert i\alpha\rangle\langle \overline{i}\alpha\vert+\frac{1}{2}\left(J_{\tau\perp}+J_{M2}\right)c^{\dagger}_{\overline{i}\alpha}c_{i\alpha}\vert i\alpha\rangle\langle \overline{i}\alpha\vert=J_{\tau\perp}c^{\dagger}_{\overline{i}\alpha}c_{i\alpha}\vert i\alpha\rangle\langle \overline{i}\alpha\vert\\
\end{aligned}
\end{equation}
since $J_{\sigma\perp}=J_{M3}$ and $J_{\tau\perp}=J_{M2}$.

Since $H_D$ is diagonal in the single impurity states, after tracing out these local states, the partition function can be expanded in $H_f$, and the results is:
\begin{equation}
Z=\sum^{\infty}_{n=0}\int^{\beta}_0d\tau_n...\int^{\tau_{i+1}}_0d\tau_i...\int^{\tau_{2}}_0d\tau_1 \sum_{m}A\left(m;\tau_n,...,\tau_1\right)\ .
\end{equation}

Here the transition amplitude is defined as:
\begin{equation}
\begin{aligned}
&A\left(m;\tau_n,...\tau_1\right)=\left(-1\right)^n\sum_{m_2,...,m_n}\int DcD\phi\exp\left[-H_{m}\left(\beta-\tau_n\right)\right]Q'\left(m,m_n\right)\times...\\
&\times\exp\left[-H_{m_{i+1}}\left(\tau_{i+1}-\tau_{i}\right)\right]Q'\left(m_{i+1},m_{i}\right)\exp\left[-H_{m_{i}}\left(\tau_{i}-\tau_{i-1}\right)\right]\times...\\
&\times\exp\left[-H_{m_2}\left(\tau_2-\tau_1\right)\right]Q'\left(m_2,m\right)\exp\left[-H_{m}\tau_1\right]\ ,
\end{aligned}
\end{equation}
where \begin{equation}
Q'\left(m_{i+1},m_i\right)=\langle m_{i+1}\vert H_f \vert m_i\rangle,
\end{equation}
 which can be separated as:
\begin{equation}
\langle m\vert H_f \vert n\rangle=y'_{m,n}O'\left(m,n\right)
\end{equation}
with 
\begin{equation}
\begin{aligned}\\
%&y'_{i\alpha,i\alpha}=\frac{1}{4}\left(J_{\sigma z}+J_{\tau z}+J_{M4}\right)\\
&y'_{i\alpha,\overline{i}\overline{\alpha}}=J_{M1}\ ,\\
&y'_{i\alpha,i\overline{\alpha}}=\frac{1}{2}\left(J_{\sigma\perp}+J_{M3}\right)=J_{\sigma\perp}\ ,\\
&y'_{i\alpha,\overline{i}\alpha}=\frac{1}{2}\left(J_{\tau\perp}+J_{M2}\right)=J_{\tau\perp}\ ,\\
%&O'_{i\alpha,i\alpha}=c^{\dagger}_{i\alpha}c_{i\alpha}\\
&O'_{i\alpha,\overline{i}\overline{\alpha}}=c^{\dagger}_{\overline{i}\overline{\alpha}}c_{i\alpha}\ ,\\
&O'_{i\alpha,i\overline{\alpha}}=c^{\dagger}_{i\overline{\alpha}}c_{i\alpha}\ ,\\
&O'_{i\alpha,i\overline{\alpha}}=c^{\dagger}_{\overline{i}\alpha}c_{i\alpha}\ .\\
\end{aligned}
\end{equation}

Now we can trace out the conduction electron by using the bosonization technique. For our single impurity problem, we only need to consider the s-wave component:
 \begin{equation}
c_{i\alpha}\left(x\right)=\frac{1}{\sqrt{2\pi a}}e^{-i\theta_{i\alpha}\left(x\right)}\ .
\end{equation}

The projected Hamiltonian thus transforms into:
\begin{equation}
H_{m}=H_{c}+H_{\phi_{\sigma }}+H_{\phi_{\tau }}+E'_{m}+\sum_{n}\frac{\delta^{n}_{m}}{\pi\rho_0}\left(\frac{d\theta_{n}\left(x\right)}{dx}\right)+\sum_{q}F^{
m}_{\sigma}\left(\phi_{\sigma z}+\phi^{\dagger}_{\sigma z,-q}\right)+\sum_{q}F^{
m}_{\tau}\left(\phi_{\tau z,q}+\phi^{\dagger}_{\tau z,-q}\right)\ ,
\end{equation}
where $E'_{m}=E_{m}+\Delta E_{m}$, $\rho_0$ is the bare conduction electron density of state, and $\delta^{j\beta}_{i\alpha}$ is the phase shift from the scattering potential:
\begin{equation}\label{phaseshift}
\begin{aligned}
&\delta^{i\alpha}_{i\alpha}=\tan^{-1}{\left(\pi\rho_0V^{i\alpha}_{i\alpha}\right)}=\tan^{-1}{\left[\frac{\pi\rho_0}{4}\left(J_{\sigma z}+J_{\tau z}+J_{M4}\right)\right]}\ ,\\
&\delta^{i\overline{\alpha}}_{i\alpha}=\tan^{-1}{\left(\pi\rho_0V^{i\overline{\alpha}}_{i\alpha}\right)}=\tan^{-1}{\left[\frac{\pi\rho_0}{4}\left(J_{\tau z}-J_{\sigma z}-J_{M4}\right)\right]}\ ,\\
&\delta^{\overline{i}\alpha}_{i\alpha}=\tan^{-1}{\left(\pi\rho_0V^{\overline{i}\alpha}_{i\alpha}\right)}=\tan^{-1}{\left[\frac{\pi\rho_0}{4}\left(J_{\sigma z}-J_{\tau z}-J_{M4}\right)\right]}\ ,\\
&\delta^{\overline{i}\overline{\alpha}}_{i\alpha}=\tan^{-1}{\left(\pi\rho_0V^{\overline{i}\overline{\alpha}}_{i\alpha}\right)}=\tan^{-1}{\left[-\frac{\pi\rho_0}{4}\left(J_{\sigma z}+J_{\tau z}-J_{M4}\right)\right]}\ .\\
\end{aligned}
\end{equation}

The history dependent potential is treated then through introducing a canonical transformation at each imaginary time:
\begin{equation}
U_{\delta}=\exp\left(i\frac{\delta}{\pi}\theta\right)\ .
\end{equation}
The potential after the canonical transformation is time-independent because of the property:

\begin{equation}
U^{\dagger}_{\delta}H_{c}U_{\delta}=H_c+\frac{\delta}{\pi\rho_0}\frac{d\theta}{dx}\ .
\end{equation}

We also introduce a similar canonical transformation to the bosonic degree of freedom,
\begin{equation}
\begin{aligned}
&U_{W_{\sigma,m}}=\exp\left(\sum_{q}\frac{F^{m}_{\sigma}}{W_{q}}\left(\phi_{\sigma z,q}-\phi^{\dagger}_{\sigma z,-q}\right)\right)\ ,\\
&U_{W_{\tau,m}}=\exp\left(\sum_{q}\frac{F^{m}_{\tau}}{W_{q}}\left(\phi_{\tau z,q}-\phi^{\dagger}_{\tau z,-q}\right)\right)\\
\end{aligned}
\end{equation}
with the property:
\begin{equation}
\begin{aligned}
&U^{\dagger}_{W_{\sigma,m}}H_{\phi_{\sigma}}U_{W_{\sigma,m}}=H_{\phi_{\sigma}}+\sum_{q}F^{
m}_{\sigma}\left(\phi_{\sigma z,q}+\phi^{\dagger}_{\sigma z,-q}\right)\ ,\\
&U^{\dagger}_{W_{\tau,m}}H_{\phi_{\tau}}U_{W_{\tau,m}}=H_{\phi_{\tau}}+\sum_{q}F^{
m}_{\tau}\left(\phi_{\tau z,q}+\phi^{\dagger}_{\tau z,-q}\right)\ .
\end{aligned}
\end{equation}

The transition amplitude now reduce to:
\begin{equation}
\begin{aligned}
&A\left(m;\tau_n,...,\tau_1\right)=Z_c\sum_{m_{n+1}=\alpha_1=m,m_2,...m_{n-1}}y'_{m_{n+1},\alpha_n}...y'_{m_{i+1},m_i}...y'_{m_{2},m_1}\\
&\times\exp\left[-E'_{m}\left(\tau_1-\tau_n\right)-\sum^{n-1}_{i=2}E'_{m_{i+1}}\left(\tau_{i+1}-\tau_i\right)\right]\\
&\times\langle O\left(m_{n+1},m_n\right)\left(\tau_n\right)...O\left(m_{i+1},m_i\right)\left(\tau_i\right)...O\left(m_{2},m_1\right)\left(\tau_1\right) \rangle\\
&\times\langle B_{\sigma}\left(m_{n+1},m_n\right)\left(\tau_n\right)...B_{\tau}\left(m_{i+1},m_i\right)\left(\tau_i\right)...B_{\sigma}\left(m_{2},m_1\right)\left(\tau_1\right) \rangle\\
&\times\langle B_{\tau}\left(m_{n+1},m_n\right)\left(\tau_n\right)...B_{\tau}\left(m_{i+1},m_i\right)\left(\tau_i\right)...B_{\tau}\left(m_{2},m_1\right)\left(\tau_1\right) \rangle\ .
\end{aligned}
\end{equation}

Here, for the bosonic part
\begin{equation}
\begin{aligned}
&B_{\sigma}\left(m_{i+1},M_i\right)\left(\tau_i\right)=U_{W_{\sigma,m_{i+1}}}U^{\dagger}_{W_{\sigma,m_{i}}}\left(\tau_i\right)\ ,\\
&B_{\tau}\left(m_{i+1},m_i\right)\left(\tau_i\right)=U_{W_{\tau,m_{i+1}}}U^{\dagger}_{W_{\tau,m_{i}}}\left(\tau_i\right)\ ,
\end{aligned}
\end{equation}
 the correlation function can be reduced into
\begin{equation}
\begin{aligned}
&\langle B_{\sigma}\left(m_{n+1},m_n\right)\left(\tau_n\right)...B_{\sigma}\left(m_{2},m_1\right)\left(\tau_1\right) \rangle=U_{W_{\sigma,m_{i+1}}}U^{\dagger}_{W_{\sigma,m_{i}}}\left(\tau_i\right)\\
&=\langle\prod_{i}\exp\left(-\sum_{q}\frac{F_{\sigma z}^{m_{i+1}m_{i}}}{W_q}\left(\phi_{\sigma z, q}-\phi^{\dagger}_{\sigma z,-q}\right)\left(\tau_i\right)\right) \rangle\\
&=\langle\exp\left(\sum_{ij}C_{\sigma}\left(\tau_i-\tau_j\right)\exp\left(\Delta_E\right)\right) \rangle
\end{aligned},
\end{equation}
and similarly
\begin{equation}
\begin{aligned}
&\langle B_{\tau}\left(m_{n+1},m_n\right)\left(\tau_n\right)...B_{\tau}\left(m_{2},m_1\right)\left(\tau_1\right) \rangle\\
&=U_{W_{\tau,m_{i+1}}}U^{\dagger}_{W_{\tau,m_{i}}}\left(\tau_i\right)=\langle\prod_{i}\exp\left(-\sum_{q}\frac{F_{\tau z}^{m_{i+1}m_{i}}}{W_q}\left(\phi_{\tau z, q}-\phi^{\dagger}_{\tau z,-q}\right)\left(\tau_i\right)\right) \rangle\\
&=\langle\exp\left(\sum_{ij}C_{\tau}\left(\tau_i-\tau_j\right)\exp\left(\Delta_E\right)\right) \rangle\ ,
\end{aligned}
\end{equation}
where 
\begin{equation}
\begin{aligned}
&F^{m_{i+1}m_i}_{\sigma }=F^{m_{i+1}}_{\sigma }-F^{m_i}_{\sigma }\ ,\\
&C_{\sigma}\left(\tau_i-\tau_j\right)=\sum_q\frac{F_{\sigma}^{m_{i+1}m_{i}}F_{\sigma}^{m_{j+1}m_{j}}}{W^2_q}\exp\left(-W_q\left(\tau_j-\tau_i\right)\right)
\end{aligned}
\end{equation}
and
\begin{equation}
\begin{aligned}
&F^{m_{i+1}m_i}_{\tau }=F^{m_{i+1}}_{\tau  }-F^{m_i}_{\tau  }\ ,\\
&C_{\tau }\left(\tau_i-\tau_j\right)=\sum_q\frac{F_{\tau }^{m_{i+1}m_{i}}F_{\tau}^{m_{j+1}m_{j}}}{W^2_q}\exp\left(-W_q\left(\tau_j-\tau_i\right)\right)
\end{aligned}
\end{equation}
with
\begin{equation}
\sum_{q}\exp\left(-W_q\tau\right)=\frac{K_{0}}{\tau^{2-\epsilon}}\ .
\end{equation}

On the other hand,  for the conduction electron part
\begin{equation}
\begin{aligned}
&O\left(m_{i+1},m_i\right)\left(\tau_i\right)=\exp\left(H_c\tau_i\right)O\left(m_{i+1},m_i\right)\exp\left(-H_c\tau_i\right)\ .
\end{aligned}
\end{equation}

Here, 
\begin{equation}
\begin{aligned}
&O\left(m_{i+1},m_i\right)=\left(\prod_{n} U_{\delta^{n}_{m_{i+1}}}\right)O' \left(m,m_i\right)\left(\prod_{n} U^{\dagger}_{\delta^{n}_{n_{i}}}\right),
\end{aligned}
\end{equation}
and for different channels, they are:
\begin{equation}\label{o1}
\begin{aligned}
%&O\left(i\alpha,i\alpha\right)=\prod_{j\beta}U^{j\beta}_{i\alpha}c^{\dagger}_{i\alpha}c_{i\alpha}\prod_{j\beta}U^{\dagger j\beta}_{i\alpha}=1\\
&O\left(i\alpha,\overline{i}\overline{\alpha}\right)=\prod_{j\beta}U^{j\beta}_{i\alpha}c^{\dagger}_{\overline{i}\overline{\alpha}}c_{i\alpha}\prod_{j\beta}U^{\dagger j\beta}_{\overline{i}\overline{\alpha}}\\
&=\exp\left[\left(\frac{\delta^{i\alpha}_{i\alpha}}{\pi}-\frac{\delta^{i\alpha}_{\overline{i}\overline{\alpha}}}{\pi}-1\right)\theta_{i\alpha}+\left(\frac{\delta^{i\overline{\alpha}}_{i\alpha}}{\pi}-\frac{\delta^{i\overline{\alpha}}_{\overline{i}\overline{\alpha}}}{\pi}\right)\theta_{i\overline{\alpha}}+\left(\frac{\delta^{\overline{i}\alpha}_{i\alpha}}{\pi}-\frac{\delta^{\overline{i}\alpha}_{\overline{i}\overline{\alpha}}}{\pi}\right)\theta_{\overline{i}\alpha}+\left(\frac{\delta^{\overline{i}\overline{\alpha}}_{i\alpha}}{\pi}-\frac{\delta^{\overline{i}\overline{\alpha}}_{\overline{i}\overline{\alpha}}}{\pi}+1\right)\theta_{\overline{i}\overline{\alpha}}\right]\ ,\\
&O\left(i\alpha,i\overline{\alpha}\right)=\prod_{j\beta}U^{j\beta}_{i\alpha}c^{\dagger}_{i\overline{\alpha}}c_{i\alpha}\prod_{j\beta}U^{\dagger j\beta}_{i\overline{\alpha}}\\
&=\exp\left[\left(\frac{\delta^{i\alpha}_{i\alpha}}{\pi}-\frac{\delta^{i\alpha}_{i\overline{\alpha}}}{\pi}-1\right)\theta_{i\alpha}+\left(\frac{\delta^{i\overline{\alpha}}_{i\alpha}}{\pi}-\frac{\delta^{i\overline{\alpha}}_{i\overline{\alpha}}}{\pi}+1\right)\theta_{i\overline{\alpha}}+\left(\frac{\delta^{\overline{i}\alpha}_{i\alpha}}{\pi}-\frac{\delta^{\overline{i}\alpha}_{i\overline{\alpha}}}{\pi}\right)\theta_{\overline{i}\alpha}+\left(\frac{\delta^{\overline{i}\overline{\alpha}}_{i\alpha}}{\pi}-\frac{\delta^{\overline{i}\overline{\alpha}}_{i\overline{\alpha}}}{\pi}\right)\theta_{\overline{i}\overline{\alpha}}\right]\ ,\\
&O\left(i\alpha,\overline{i}\alpha\right)=\prod_{j\beta}U^{j\beta}_{i\alpha}c^{\dagger}_{\overline{i}\alpha}c_{i\alpha}\prod_{j\beta}U^{\dagger j\beta}_{\overline{i}\alpha}\\
&=\exp\left[\left(\frac{\delta^{i\alpha}_{i\alpha}}{\pi}-\frac{\delta^{i\alpha}_{\overline{i}\alpha}}{\pi}-1\right)\theta_{i\alpha}+\left(\frac{\delta^{i\overline{\alpha}}_{i\alpha}}{\pi}-\frac{\delta^{i\overline{\alpha}}_{\overline{i}\alpha}}{\pi}\right)\theta_{i\overline{\alpha}}+\left(\frac{\delta^{\overline{i}\alpha}_{i\alpha}}{\pi}-\frac{\delta^{\overline{i}\alpha}_{\overline{i}\alpha}}{\pi}+1\right)\theta_{\overline{i}\alpha}+\left(\frac{\delta^{\overline{i}\overline{\alpha}}_{i\alpha}}{\pi}-\frac{\delta^{\overline{i}\overline{\alpha}}_{\overline{i}\alpha}}{\pi}\right)\theta_{\overline{i}\overline{\alpha}}\right]\ .\\
\end{aligned}
\end{equation}

We can rewrite these term as:
\begin{equation}
\begin{aligned}
&O\left(m,n\right)=\exp\left[i\sum_{r}e^{r}_{mn}\theta_{r}\right]\ .\\
\end{aligned}
\end{equation}

After all of these, the partition function is mapped into: 
\begin{equation}
\begin{aligned}
&\frac{Z}{Z_0}=\sum^{\infty}_{n=0}\sum_{m_{n+1}=m_1=m,m_2,...m_{n-1}}\int^{\beta-\xi_0}_{\xi_0}\frac{d\tau_n}{\xi_0}...\int^{\tau_{i+1}-\xi_0}_{\xi_0}\frac{d\tau_i}{\xi_0}...\int^{\tau_2-\xi_0}_{\xi_0}\frac{d\tau_1}{\xi_0}\exp\left[-S\left(\tau_1,...,\tau_n\right)\right]\\
\end{aligned}
\end{equation}
with a Coulomb gas type action:
\begin{equation}\label{coulombgas}
\begin{aligned}
&S\left(\tau_1,...,\tau_n\right)=-\sum_{i}\ln {y_{m_i, m_{i+1}}}+\sum_{i}h_{m_{i+1}}\frac{\tau_{i+1}-\tau_i}{\xi_0}\\
&+\sum_{i<j}\left[K_{m_i,m_j}+K_{m_{i+1},m_{j+1}}-K_{m_i,m_{j+1}}-K_{m_{i+1},m_j}\right]\ln{\frac{\tau_j-\tau_i}{\xi_0}}\\
&-\sum_{i<j}\left[M^{\sigma}_{m_i,m_j}+M^{\sigma}_{m_{i+1},m_{j+1}}-M^{\sigma}_{m_i,m_{j+1}}-M^{\sigma}_{m_{i+1},m_j}\right]\left[\left(\frac{\tau_j-\tau_i}{\xi_0}\right)^{\epsilon}-1\right]\\
&-\sum_{i<j}\left[M^{\tau}_{m_i,m_j}+M^{\tau}_{m_{i+1},m_{j+1}}-M^{\tau}_{m_i,m_{j+1}}-M^{\tau}_{m_{i+1},m_j}\right]\left[\left(\frac{\tau_j-\tau_i}{\xi_0}\right)^{\epsilon}-1\right]\ ,\\
\end{aligned}
\end{equation}
where $h_{m}\propto E'_m$, ${\xi_0}$ is the ultraviolet cutoff, and
\begin{equation}\label{stifffuga}
\begin{aligned}
&y_{m,n}=y'_{m,n}{\xi_0}\ ,\\
&K_{m,n}=-\frac{1}{2}\sum_{r}\left(e^{r}_{m n}\right)^2\ ,\\
&M^{\sigma}_{m,n}= -\frac{1}{2}\sum_{q}\left(F^{m n}_{\sigma }\right)^2\ ,\\
&M^{\tau}_{m,n}=-\frac{1}{2}\sum_{q}\left(F^{m n}_{\tau }\right)^2\ .\\
\end{aligned}
\end{equation}

By following these definitions, for the Bose-Fermi Kondo model (Eq.\,\ref{apSU4RGmodel}) the non-vanishing fugacity $y_{m,n}$ and stiffness $K_{m,n}$, $M^{\sigma}_{m,n}$ and $M^{\tau}_{m,n}$ are:
\begin{equation}
\begin{aligned}
&y_{i\alpha,\overline{i}\overline{\alpha}}\equiv y_1=\xi_0J_{M1}\ ,\\
&y_{i\alpha,i\overline{\alpha}}\equiv y_2=\xi_0J_{\sigma\perp}\ ,\\
&y_{i\alpha,\overline{i}\alpha}\equiv y_3=\xi_0J_{\tau\perp}\ ,\\
&K_{i\alpha,\overline{i}\overline{\alpha}}\equiv -K_1=-f_1\left(J_{\sigma z},J_{\tau z}, J_{M4}\right)\ ,\\
&K_{i\alpha,i\overline{\alpha}}\equiv -K_2=-f_2\left(J_{\sigma z},J_{\tau z}, J_{M4}\right)\ ,\\
&K_{i\alpha,\overline{i}\alpha}\equiv -K_3=-f_3\left(J_{\sigma z},J_{\tau z}, J_{M4}\right)\ ,\\
&M^{\sigma}_{i\alpha,\overline{i}\overline{\alpha}}=M^{\sigma}_{i\alpha,i\overline{\alpha}}\equiv -M^{\sigma}=-\Gamma\left(\epsilon\right) g^2_{\sigma z}\ ,\\
&M^{\tau}_{i\alpha,\overline{i}\overline{\alpha}}=M^{\tau}_{i\alpha,\overline{i}\alpha}\equiv -M^{\tau}=- \Gamma\left(\epsilon\right)  g^2_{\tau z}\ ,\\
\end{aligned}
\end{equation}
where $\Gamma\left(\epsilon\right)$ is a $\epsilon$ dependent O(1) constant. The explicit expression of $K_{1,2,3}$ is complicated but unnecessary, and can be derived from Eq.\,\ref{o1}, where the phase shifts are known in Eq.\,\ref{phaseshift}.  The only few things that matter are that they depend only on indices-preserving coupling $J_{\sigma z, \tau z, M4}$, and the range of their bare value is $f_{1,2,3}\left(J_{\sigma z},J_{\tau z}, J_{M4}\right)\in\left(0,3\right)$.

\begin{figure}[htbp]
   \centering
   \includegraphics[scale=1]{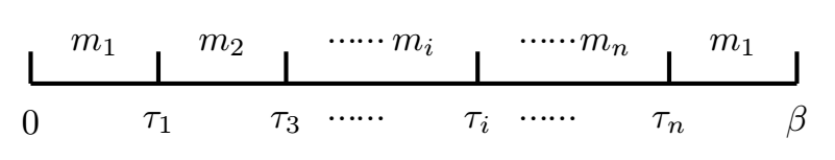}
   \caption{   \label{fig:sequence}
Hopping sequences of the local states along the imaginary time axis. Here $\tau_i$, for $i=1,...,n$, labels the imaginary time at which the local state hops from $\vert m_i\rangle$ to $\vert m_{i+1}\rangle$. 
 }
\end{figure}

The Coulomb gas action (Eq.\,\ref{coulombgas}) is a summation over all possible histories of the local degrees of freedom which fluctuate between $n+1$ local states $ \vert m \rangle$. Each history, labeled by $\lbrace m_1,\ldots,m_n;\tau_1,\ldots\tau_n \rbrace$ , is a sequence of the transition between the local states from $m_1$ through $m_n$ taking place at the time $\tau_1<\ldots<\tau_n$, as illustrated in 
Fig.\,\ref{fig:sequence}.
(The periodic boundary condition has been imposed.)
The action (Eq.\,\ref{coulombgas}) gives the statistical weight of such a history. We can interpret 
Eq.\,\ref{coulombgas} in terms of a partition function of a plasma of kinks with interactions. It has multiple components of ``stiffness" $K_{m,n}$ and  $M^{\tau/\sigma}_{m,n}$  and ``fugacity" $y_{m,n}$,
as defined in (Eq.\,\ref{stifffuga}). 
For such 
a
Coulomb gas action (Eq.\,\ref{coulombgas}), one can perform the RG calculation by integrating out the degrees of freedom within the cutoff shell $\left[\xi_0,\xi_0+\xi_0dl\right]$ (Refs.\,
\citenum{Qimiao1996,Smith1999,Chakravarty1982,Cardy1981,Qimiao1993}). 
In the absence of the $M^{\tau/\sigma}_{m,n}$ terms,
 the RG beta functions have been derived in Ref.\,\citenum{Cardy1981}. 
 With the $M^{\tau/\sigma}_{m,n}$ terms, the same technique is still applicable \cite{Smith1999,Smith2000}, and the final beta functions are:
\begin{equation}\label{BFKBeta}
\begin{aligned}
&\frac{dy_{1}}{dl}=\left(1-K_{1}-M^{\sigma}-M^{\tau}\right)y_{1}+2y_2y_3\ ,\\
&\frac{dy_{2}}{dl}=\left(1-K_{2}-M^{\sigma}\right)y_{2}+2y_1y_3\ ,\\
&\frac{dy_{3}}{dl}=\left(1-K_{3}-M^{\tau}\right)y_{3}+2y_1y_2\ ,\\
&\frac{dK_{1}}{dl}=-2y^2_1\left(2K_1\right)-2y^2_2\left(K_1+K_2-K_3\right)-2y^2_3\left(K_1+K_3-K_2\right)\ ,\\
&\frac{dK_{2}}{dl}=-2y^2_1\left(K_2+K_1-K_3\right)-2y^2_2\left(2K_2\right)-2y^2_3\left(K_2+K_3-K_1\right)\ ,\\  
&\frac{dK_{3}}{dl}=-2y^2_1\left(K_3+K_1-K_2\right)-2y^2_2\left(K_3+K_2-K_1\right)-2y^2_3\left(2K_3\right)\ ,\\      
&\frac{dM^{\sigma}}{dl}=\left(\epsilon -4y^2_1-4y^2_2\right)M^{\sigma}\ ,\\  
&\frac{dM^{\tau}}{dl}=\left(\epsilon -4y^2_1-4y^2_3\right)M^{\tau}\ .\\     
\end{aligned}
\end{equation}

%\section{
\noindent
\subsection*{D.~RG analysis and the generic phase diagram}
%\label{appen:RGphase}

In this section, we give a detailed RG analysis of the beta functions (Eq.\,\ref{BFKBeta}). We will identify the fixed points of the  
beta functions (Eq.\,\ref{BFKBeta}) by using $\epsilon$ as the control parameter. The relative stability of these fixed points are analysed through the eigenvalues and eigenvectors of the matrix:
\begin{equation}\label{matrix}
W=\begin{pmatrix}
\frac{\partial\beta_{y_1}}{\partial y_1}&\frac{\partial\beta_{y_1}}{\partial y_2}&\frac{\partial\beta_{y_1}}{\partial y_3}&\frac{\partial\beta_{y_1}}{\partial K_1}&\frac{\partial\beta_{y_1}}{\partial K_2}&\frac{\partial\beta_{y_1}}{\partial K_3}&\frac{\partial\beta_{y_1}}{\partial M^{\sigma}}&\frac{\partial\beta_{y_1}}{\partial M^{\tau}}\\
\frac{\partial\beta_{y_2}}{\partial y_1}&\frac{\partial\beta_{y_2}}{\partial y_2}&\frac{\partial\beta_{y_2}}{\partial y_3}&\frac{\partial\beta_{y_2}}{\partial K_1}&\frac{\partial\beta_{y_2}}{\partial K_2}&\frac{\partial\beta_{y_2}}{\partial K_3}&\frac{\partial\beta_{y_2}}{\partial M^{\sigma}}&\frac{\partial\beta_{y_2}}{\partial M^{\tau}}\\
\frac{\partial\beta_{y_3}}{\partial y_1}&\frac{\partial\beta_{y_3}}{\partial y_2}&\frac{\partial\beta_{y_3}}{\partial y_3}&\frac{\partial\beta_{y_3}}{\partial K_1}&\frac{\partial\beta_{y_3}}{\partial K_2}&\frac{\partial\beta_{y_3}}{\partial K_3}&\frac{\partial\beta_{y_3}}{\partial M^{\sigma}}&\frac{\partial\beta_{y_3}}{\partial M^{\tau}}\\
\frac{\partial\beta_{K_1}}{\partial y_1}&\frac{\partial\beta_{K_1}}{\partial y_2}&\frac{\partial\beta_{K_1}}{\partial y_3}&\frac{\partial\beta_{K_1}}{\partial K_1}&\frac{\partial\beta_{K_1}}{\partial K_2}&\frac{\partial\beta_{K_1}}{\partial K_3}&\frac{\partial\beta_{K_1}}{\partial M^{\sigma}}&\frac{\partial\beta_{K_1}}{\partial M^{\tau}}\\
\frac{\partial\beta_{K_2}}{\partial y_1}&\frac{\partial\beta_{K_2}}{\partial y_2}&\frac{\partial\beta_{K_2}}{\partial y_3}&\frac{\partial\beta_{K_2}}{\partial K_1}&\frac{\partial\beta_{K_2}}{\partial K_2}&\frac{\partial\beta_{K_2}}{\partial K_3}&\frac{\partial\beta_{K_2}}{\partial M^{\sigma}}&\frac{\partial\beta_{K_2}}{\partial M^{\tau}}\\
\frac{\partial\beta_{K_3}}{\partial y_1}&\frac{\partial\beta_{K_3}}{\partial y_2}&\frac{\partial\beta_{K_3}}{\partial y_3}&\frac{\partial\beta_{K_3}}{\partial K_1}&\frac{\partial\beta_{K_3}}{\partial K_2}&\frac{\partial\beta_{K_3}}{\partial K_3}&\frac{\partial\beta_{K_3}}{\partial M^{\sigma}}&\frac{\partial\beta_{K_3}}{\partial M^{\tau}}\\
\frac{\partial\beta_{M^{\sigma}}}{\partial y_1}&\frac{\partial\beta_{M^{\sigma}}}{\partial y_2}&\frac{\partial\beta_{M^{\sigma}}}{\partial y_3}&\frac{\partial\beta_{M^{\sigma}}}{\partial K_1}&\frac{\partial\beta_{M^{\sigma}}}{\partial K_2}&\frac{\partial\beta_{M^{\sigma}}}{\partial K_3}&\frac{\partial\beta_{M^{\sigma}}}{\partial M^{\sigma}}&\frac{\partial\beta_{M^{\sigma}}}{\partial M^{\tau}}\\
\frac{\partial\beta_{M^{\tau}}}{\partial y_1}&\frac{\partial\beta_{M^{\tau}}}{\partial y_2}&\frac{\partial\beta_{M^{\tau}}}{\partial y}&\frac{\partial\beta_{M^{\tau}}}{\partial K_1}&\frac{\partial\beta_{M^{\tau}_3}}{\partial K_2}&\frac{\partial\beta_{M^{\tau}}}{\partial K_3}&\frac{\partial\beta_{M^{\tau}}}{\partial M^{\sigma}}&\frac{\partial\beta_{M^{\tau}}}{\partial M^{\tau}}\\
\end{pmatrix}\ . \\ 
\end{equation},

We will also illustrate the generic phase diagram 
Fig.\,\ref{fig:su4RGre}(a)
based on our RG analysis.

%\subsection{RG analysis }
\noindent{\bf RG analysis.}
In the $\epsilon$ expansion, the $\epsilon$ serve as a small control parameter.
We will express the fixed point in term of the $\epsilon$ up to the leading order $\sqrt{\epsilon}$. 
By solving the zeros of the beta functions (Eq.\,\ref{BFKBeta}),  the fixed points can be identified:
\begin{equation}\label{fixedp1}
\begin{aligned}
&\textbf{R1}:\;
y_1=0,\:y_2=y_3=\frac{\sqrt{\epsilon}}{2},\:K_1=K_2=K_3=0,\: M^{\sigma}=M^{\tau}=1\ ,\\
&\textbf{R2}:\;
y_1=\frac{\sqrt{\epsilon}}{2},\:y_2=0,\:y_3=0,\:K_2=K_3,\: K_1=0\ ,\: M^{\sigma}+M^{\tau}=1,\\
\end{aligned}
\end{equation}
where the RG trajectory around \textbf{R2} can flow toward \textbf{R1}. Other fixed points includes
\begin{equation}\label{fixedp2}
\begin{aligned}
&\textbf{E1}:\;
y_1=0,\:y_2=\frac{\sqrt{\epsilon}}{2},\:y_3=0,\:K_1=K_3,\: K_2=0,\: M^{\sigma}=1\ ,\: M^{\tau}=0\ ,\\
\end{aligned}
\end{equation}
and
\begin{equation}\label{fixedp3}
\begin{aligned}
&\textbf{E2}:\;
y_1=0,\:y_2=0,\:y_3=\frac{\sqrt{\epsilon}}{2},\:K_1=K_2,\: K_3=0,\: M^{\sigma}=0,\: M^{\tau}=1 \ ,\\
\end{aligned}
\end{equation}
and both of the fixed points \textbf{E1} and \textbf{E2} are unstable and the RG trajectory around them can flow toward \textbf{R1} and \textbf{R2}. Finally, there is a unstable fixed points
\begin{equation}\label{fixedp4}
\begin{aligned}
&\textbf{E3}:\;
y_1=y_2=y_3=0,\: M^{\sigma}=M^{\tau}=0\ .
\end{aligned}
\end{equation}
We will ignore this fixed point in the following, since it is the most unstable fixed points.\\

Among the fixed points listed in the Eqs.\,\ref{fixedp1}-\ref{fixedp4}, the fixed point \textbf{R1} is the most stable one. However, the fixed point \textbf{R1} is actually still not a generic critical point, since there are two relevant directions $\vec{v}_{1,2}$ around it. The first one is:
\begin{equation}
 \vec{v}_1=\frac{1}{2\sqrt{2}}\hat{y}_2-\frac{1}{2\sqrt{2}}\hat{y}_3+\hat{M}^{\sigma}-\hat{M}^{\tau}
\end{equation} 
which has the associated eigenvalue scaling dimension $\sqrt{2\epsilon}$ and can flow toward either the orbital KS fixed point:
\[
\textbf{K1}:\;
y_2\rightarrow \infty,\: y_1=y_3=0,\:K_1=K_2=K_3,\: M^{\sigma}=0,\:M^{\tau}\rightarrow \infty
\]
or spin KS fixed point:
\[
\textbf{K2}:\;
y_3\rightarrow \infty,\: y_1=y_2=0,\:K_1=K_2=K_3=0,\: M^{\sigma}\rightarrow\infty,\:M^{\tau}=0\ .
\]

On the other hand, the second relevant direction is(we express each non-vanishing coefficients up to the leading order $\sqrt{\epsilon}$):
\begin{equation}
 \vec{v}_2=\frac{\sqrt{\epsilon}}{2}\hat{y}_1+\frac{1}{2\sqrt{2}}\hat{y}_2+\frac{1}{2\sqrt{2}}\hat{y}_3-\hat{M}^{\sigma}-\hat{M}^{\tau}
\end{equation} 
which has associated  scaling dimension $\sqrt{2\epsilon}$(up to the leading order $\sqrt{\epsilon}$)
and flows toward either the strong coupling SU(4) Kondo-screened (KS) fixed point
\[
\textbf{K3}:\;
y_1,y_2,y_3\rightarrow \infty,\:K_1=K_2=K_3=0,\: M^{\sigma} =M^{\tau}=0
\]
or the spin and orbital Kondo-destroyed(KD) phase
\[
\textbf{G}:\;
y_1=y_2=y_3=0,\: M^{\sigma}\rightarrow \infty,\: M^{\tau}\rightarrow \infty\ .
\]

The stability of the strong coupling fixed points \textbf{K1}, \textbf{K2}, \textbf{K3}, and \textbf{G}
can be studied through the stability matrix $W$ (Eq.\,\ref{matrix}). 
This analysis shows 
that
the fixed points are stable against other small perturbations and thus characterize the phase of matter. 
Accordingly, there should be other generic critical points separate these phases. Since there are two relevant directions around \textbf{R1}, there should be four generic critical points separating these phase. \\

Moreover, besides flowing toward to \textbf{R1}, \textbf{K1}, \textbf{K2} and \textbf{G}, by exploring the relevant direction around the fixed point \textbf{E1} one can also checks that the RG trajectory around it can also flow toward $M^{\tau}\rightarrow \infty$( $M^{\sigma}\rightarrow \infty$), and thus approach to:
\[
\textbf{F1}:\;
y_1=0,\:y_2=\frac{\sqrt{\epsilon}}{2},\:y_3=0,\:K_1=K_2=K_3=0,\: M^{\sigma}=1,\: M^{\tau}\rightarrow \infty\ .
\]

For the \textbf{F1}, except the beta function $dM^{\tau}/dl$, other beta functions remain zero. As a result, \textbf{F1} corresponds to a fixed point at the large $M^{\tau}$ regime. By study the nearby RG trajectory through the matrix $W$ in Eq.\,\ref{matrix}, one can conclude that fixed point \textbf{F1} actually corresponds to a generic critical point separate the spin and orbital KD phase \textbf{G} and the orbital KS phase \textbf{K1}.

Similarly the RG trajectory around \textbf{E2} can flow toward:
\[
\textbf{F2}:\;
y_1=0,\:y_2=0 ,\:y_3=\frac{\sqrt{\epsilon}}{2},\:K_1=K_2=K_3=0,\: M^{\tau}=1,\: M^{\sigma}\rightarrow \infty \ ,
\]
which is a generic critical point between the spin and orbital KD phase \textbf{G} and the orbital KS phase \textbf{K3}.\\

\begin{figure}[htbp]
   \centering
   \includegraphics[scale=0.45]{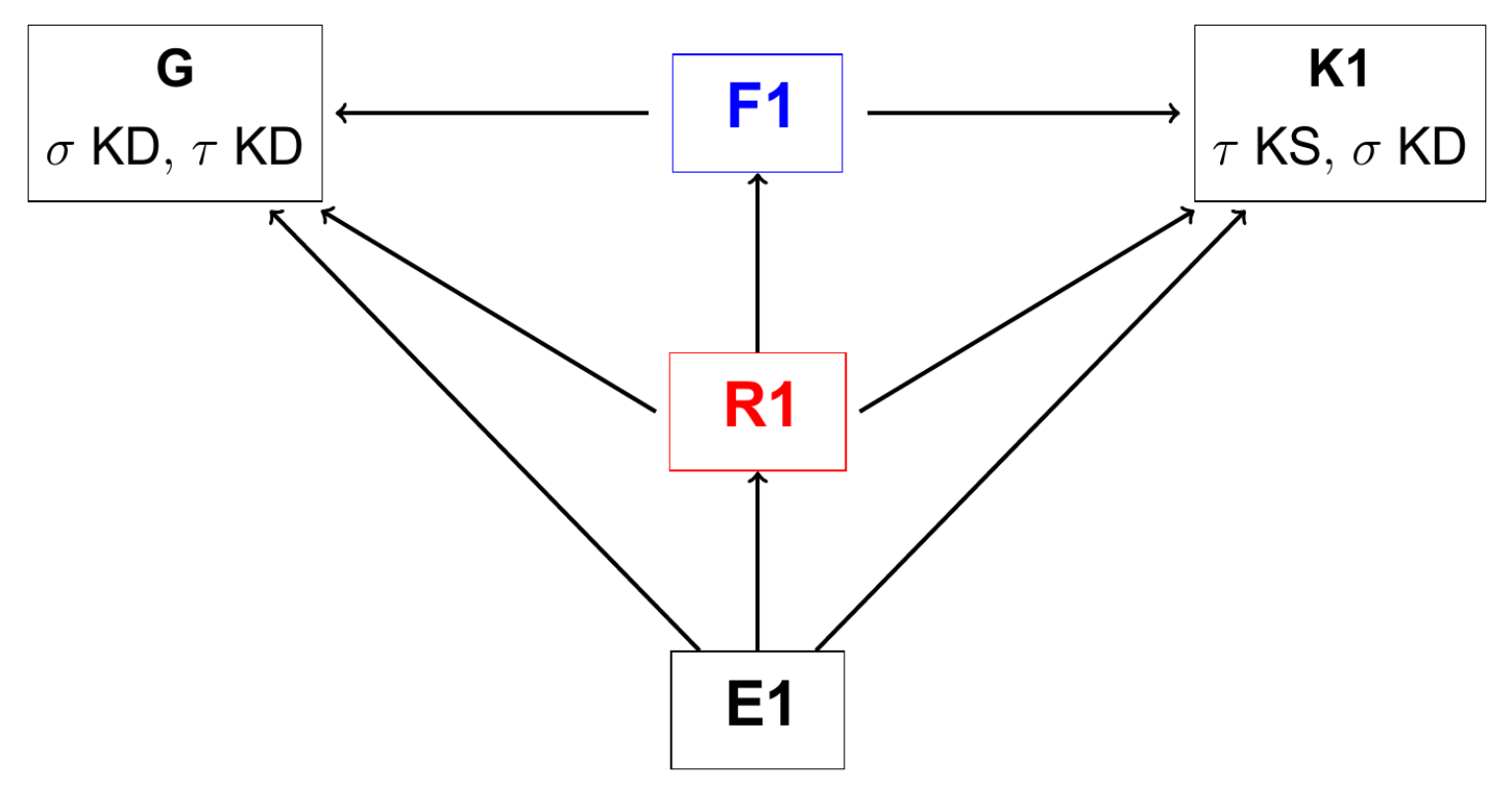}
   \caption{   \label{fig:relativeRG}
{\bf The structure of the renormalization-group flow.}
   The  relative RG flow structure among the fixed points  \textbf{R1}, \textbf{E1}, \textbf{F1}, \textbf{K1}, and \textbf{G}. 
 }
\end{figure}

The whole RG flow structure is summarized in Fig.\,\ref{fig:su4RGresultmain} of the main text, where the blue boxes are the critical points \textbf{F1}-\textbf{F2}
corresponding to the phase transitions from spin and orbital KD phase to spin or orbital KS phases.  
One can see that the spin and orbital KD phase \textbf{G} can transit to different kinds o
f strong Kondo coupling fixed points \textbf{K1}, \textbf{K2}, and \textbf{K3}.
Note that because the fixed point \textbf{R1}, denoted as the red box in
Fig.\,\ref{fig:su4RGresultmain}, is not a generic but a multi-critical point, the phase transition between 
 spin and orbital KS phase \textbf{K3} and  spin and orbital KD phase \textbf{G} should be a fine-tuned one.
 Again, since the Kondo screened fixed points \textbf{K1}, \textbf{K2}, \textbf{K3} are stable fixed points, 
 there should be some generic critical points
 (denoted as the orange boxes \textbf{X1} and \textbf{X2} in Fig.\,\ref{fig:su4RGresultmain}
 separating them, even though their exact directions is unknown in this scheme unlike the generic critical point \textbf{F1} and \textbf{F2}. 
 Based on the whole RG flow structure Fig.\,\ref{fig:su4RGresultmain}, the generic phase diagram is sumarrized 
 in 
Fig.\,\ref{fig:su4RGre}.
 Note that in Fig.\,\ref{fig:su4RGresultmain}, we neglected the fixed point \textbf{R2}, 
 \textbf{E1}, \textbf{E2} , and \textbf{E3} since these fixed points do not influence the RG structure. 
 We present the relative RG flow structure among the fixed points  \textbf{R1}, \textbf{E1}, \textbf{F1}, \textbf{K1}, 
 and \textbf{G} in Fig.\,\ref{fig:relativeRG}. \\

We emphasize that the RG flow structure Fig.\,\ref{fig:su4RGresultmain} 
is rigorously derived through the matrix $W$  (Eq.\,\ref{matrix}). However, due to the huge number of the coupling constants, it is not easy to visualize the full RG flow structure. In the following, we are going to elaborate these results in a reduced but more transparent and visible way.

\noindent
{\bf Transition to the SU(4) Kondo-screened phase.}
To illustrate the transition between the spin and orbital KD phase to the SU(4) KS phase, we choose to scan the RG flow structure by taking $g_{\sigma z}=g_{\tau z}=g$, which corresponds to the trajectory denoted as the arrow (\rom{1}) in  
Fig.\,\ref{fig:su4RGre}(a).

Since along this direction, the beta functions (Eq.\,\ref{BFKBeta}) are invariant under $\sigma\leftrightarrow \tau$, one can set $y_2=y_3=y$ and $K_2=K_3=K$.  The beta functions (Eq.\,\ref{BFKBeta}) thus can be reduced to:
\begin{equation}\label{reduceSU4apa}
\begin{aligned}
&\frac{dy_{1}}{dl}=\left(1-K_{1}-2M\right)y_{1}+2y^2\ ,\\
&\frac{dy}{dl}=\left(1-K-M\right)y+2y_1y\ ,\\
&\frac{dK_{1}}{dl}=-2y^2_1\left(2K_1\right)-4y^2\left(K_1\right)\ ,\\
&\frac{dK}{dl}=-2y^2_1\left(K_1\right)-4y^2\left(2K\right)+2y^2\left(K_1\right)\ ,\\  
&\frac{dM}{dl}=\left(\epsilon -4y^2_1-4y^2\right)M\ .\\    
\end{aligned}
\end{equation}

Note that the fugacity $y_1$ flips both spin and orbital part, while the fugacity
$y$ flips only the spin or orbital index. As a result, the beta functions of $y_1$ and $y$ involve $y^2$ and $y_1y$, respectively.

From the  beta functions {Eq.\,\ref{reduceSU4apa}}, one can see that the coupling constant $K_1$ flows to $0$ no matter the initial values, and thus the beta functions can be further reduced into:
\begin{equation}
\begin{aligned}
&\frac{dy_{1}}{dl}=\left(1-2M\right)y_{1}+2y^2\ ,\\
&\frac{dy}{dl}=\left(1-K-M\right)y+2y_1y\ ,\\
&\frac{dK}{dl}=-4y^2\left(2K\right)\ ,\\  
&\frac{dM}{dl}=\left(\epsilon -4y^2_1-4y^2\right)M\ ,\\    
\end{aligned}
\end{equation}
and again, $K\rightarrow 0$ no matter the initial values, so in the end we derive the reduced beta functions Eq.\,\ref{reduceSU4}: 
\begin{equation}\label{reduceSU4ap}
\begin{aligned}
&\frac{dy_{1}}{dl}=\left(1-2M\right)y_{1}+2y^2\ ,\\
&\frac{dy}{dl}=\left(1-M\right)y+2y_1y\ ,\\
&\frac{dM}{dl}=\left(\epsilon -4y^2_1-4y^2\right)M\ .\\    
\end{aligned}
\end{equation}\\

From these reduced beta functions (Eq.\,\ref{reduceSU4ap}), we 
identify the generic critical point $\left(y^*_1,y^*,M^*\right)=\left(\frac{-1+\sqrt{1+12\epsilon}}{12},\frac{\sqrt{-1+12\epsilon+\sqrt{1+12\epsilon}}}{6\sqrt{2}},\frac{5+\sqrt{1+12\epsilon}}{6}\right)\cong\left(0,\frac{\sqrt{\epsilon}}{2},1\right)$ up to the order $\sqrt{\epsilon}$. 
This critical point corresponds to the critical point \textbf{R1} in Fig.\,\ref{fig:su4RGresultmain}, 
and separates the spin and orbital KD phase from the SU(4) KS phase. 

\noindent
{\bf Transition to spin or orbital Kondo-screened phase.}
Here we aim to illustrate the transition between the spin and orbital KD phase and the spin or orbital KS phase. 
We firstly focus on the RG trajectory around the critical point \textbf{R1} where $g_{\sigma z}=g_{\tau z}=g^*$
 between the spin and orbital KD \textbf{G} and SU(4) KS phases \textbf{K3}. 

As mentioned, any small asymmetry between $g_{\tau z}$ and $g_{\sigma z}$ around \textbf{R1} actually is relevant in RG sense. Suppose we keep every parameters fixed but just slightly increase the coupling constant $g_{\tau z}$, that is, $g_{\tau z}> g_{\sigma z}=g^*$, then the RG trajectory will flow toward to $g_{\tau z}\rightarrow \infty$. We can then vary $g_{\sigma z}$ to explore the RG trajectory.  The corresponding trajectories in the phase diagram are denoted as the arrow (\rom{2}) in 
Fig.\,\ref{fig:su4RGre}(a).
Around this trajectory, according to the beta functions (Eq.\,\ref{BFKBeta}), 
$y_1$ and $y_3$ must both flow to $0$ and both are irrelevant since $g_{\tau z}\rightarrow \infty$. The beta functions can thus be reduced into:
\begin{equation}
\begin{aligned}
&\frac{dy_{2}}{dl}=\left(1-K_{2}-M^{\sigma}\right)y_{2}\ ,\\
&\frac{dK_{1}}{dl}=-2y^2_2\left(K_1+K_2-K_3\right)\ ,\\
&\frac{dK_{2}}{dl}=-2y^2_2\left(2K_2\right)\ ,\\  
&\frac{dK_{3}}{dl}=-2y^2_2\left(K_3+K_2-K_1\right)\ ,\\      
&\frac{dM^{\sigma}}{dl}=\left(\epsilon -4y^2_2\right)M^{\sigma}\ .\\  
\end{aligned}
\end{equation}
by which one can see that $K_2\rightarrow 0$, and again the beta functions can be further reduced into:
\begin{equation}\label{reducespinap}
\begin{aligned}
&\frac{dy_{2}}{dl}=\left(1-M^{\sigma}\right)y_{2}\ ,\\
&\frac{dK_{1}}{dl}=-2y^2_2\left(K_1-K_3\right)\ ,\\
&\frac{dK_{3}}{dl}=-2y^2_2\left(K_3-K_1\right)\ ,\\      
&\frac{dM^{\sigma}}{dl}=\left(\epsilon -4y^2_2\right)M^{\sigma}\ .\\  
\end{aligned}
\end{equation}

\begin{figure}[ht]
   \centering
   \includegraphics[scale=0.43]{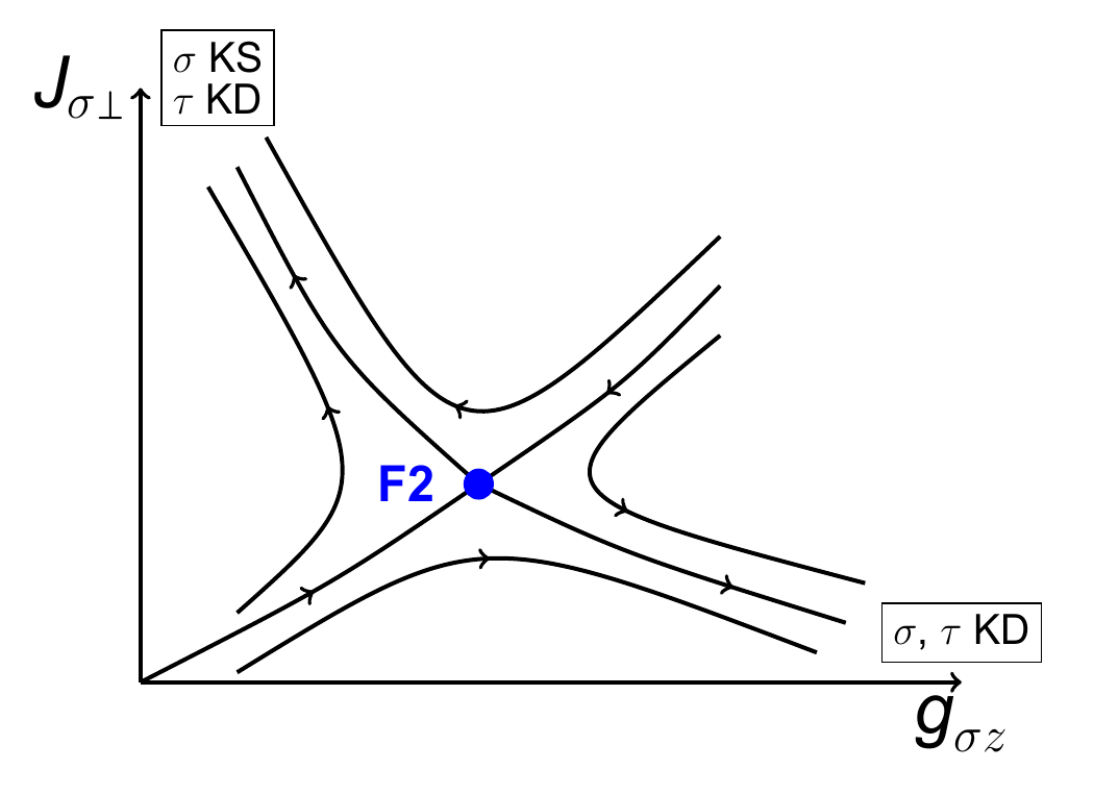}
   \caption{   \label{fig:su4RGre2}   
 {\bf A renormalization-group flow.}
   RG flow diagram of the reduced beta functions (Eq.\,\ref{reducespin}) on the $J_{\sigma\perp}-g_{\sigma z}$ plane.} 
\end{figure}

From the reduced beta functions (Eq.\,\ref{reducespinap}), one can immediately conclude that the $K_1$ and $K_3$ flow to the fixed point $K_1=K_3=k_{\tau}$, where $k_{\tau}$ is a constant.  As a result, the final reduced beta functions are indeed Eq.\,\ref{reducespin}, from which one can find a generic critical point $\left(y^*_2,M^{\sigma *}\right)=\left(\frac{\sqrt{\epsilon}}{2},1\right)$ with the scaling dimensions $\frac{1}{2}\sqrt{\epsilon}\left(\sqrt{\epsilon}+\sqrt{8+\epsilon}\right)\cong\sqrt{2\epsilon}$ (up to the order $\sqrt{\epsilon}$) that corresponds to the fixed point \textbf{F2} in Fig.\,\ref{fig:su4RGresultmain} and separates the spin and orbital KD phase from the spin KS phase. The RG flow diagram of the reduced beta functions (Eq.\,\ref{reducespin}) on the $J_{\sigma\perp}-g_{\sigma z}$ plane is shown in Fig.\,\ref{fig:su4RGre2}.

\noindent
{\bf Transition between spin or orbital KS phase and SU(4) KS phase.}
Finally, we would like to establish the transition between the spin or orbital KS phase and the SU(4) KS phase, which correspond to the trajectories \rom{3} in 
Fig.\,\ref{fig:su4RGre}(a).
As discussed in the main text, because the strong Kondo coupling fixed points \textbf{K1}, \textbf{K2}, and \textbf{K3} are stable fixed points, there should be other generics critical points, denoted as orange boxes \textbf{X1} and \textbf{X2}, separating them. 

Again, we  focus on the RG trajectory around the critical point \textbf{R1} where $g_{\sigma z}= g_{\tau z}=g^*$ between the spin and orbital KD \textbf{G} and SU(4) KS phases \textbf{K3}. If we keep every parameters fixed but just slightly decrease the coupling constant $g_{\sigma z}$, that is, $g_{\sigma z}< g_{\tau z}=g^*$, then the RG trajectory will flow toward to $g_{\sigma z}\rightarrow 0$. We can then vary $g_{\tau z}$ to explore the RG trajectory.  The corresponding trajectories in the phase diagram are denoted as the arrow (\rom{3}) in 
Fig.\,\ref{fig:su4RGre}(a).
As we will see later, the assumption that $g_{\sigma z}\rightarrow 0$ is legitimate since $g_{\sigma z}$ is generally irrelevant around $g_{\sigma z}=0$. 

However, unlike \textbf{R1} and \textbf{F1}, the real locations of the \textbf{X1} is hard to identify directly from the beta functions (Eq.\,\ref{BFKBeta}). To proceed, we exploit one more property of the critical point \textbf{R1}, that is, $y_1\sim 0$, $y_{2,3}\sim \frac{\sqrt{\epsilon}}{2}$ around \textbf{R1}.  Near the vicinity of \textbf{R1}, one can thus neglect the higher order terms of $\sqrt{\epsilon}$ in the beta functions (Eq.\,\ref{BFKBeta}).  To simplify the analysis, we also set the new variables:
\begin{equation}
\begin{aligned}
&u_3=K_1+K_2-K_3\ ,\\
&u_2=K_1+K_3-K_2\ ,\\
&u_1=K_2+K_3-K_1\ .\\ 
\end{aligned}
\end{equation}
The beta functions (Eq.\,\ref{BFKBeta}) then become:
\begin{equation}
\begin{aligned}
&\frac{dy_{1}}{dl}=\left(1-\frac{u_2}{2}-\frac{u_3}{2}-M^{\tau}\right)y_{1}\ ,\\
&\frac{dy_{2}}{dl}=\left(1-\frac{u_1}{2}-\frac{u_3}{2}\right)y_{2}\ ,\\
&\frac{dy_{3}}{dl}=\left(1-\frac{u_1}{2}-\frac{u_2}{2}-M^{\tau}\right)y_{3}\ ,\\
&\frac{du_{3}}{dl}=-4\left( y^2_1+y^2_2\right) u_3\ ,\\
&\frac{du_{2}}{dl}=-4\left( y^2_1+y^2_3\right) u_2\ ,\\  
&\frac{du_{1}}{dl}=-4\left(y^2_2+y^2_3\right)u_1\ ,\\   
&\frac{dM^{\sigma}}{dl}=\left(\epsilon -4y^2_1-4y^2_2\right)M^{\sigma}\ ,\\  
&\frac{dM^{\tau}}{dl}=\left(\epsilon -4y^2_1-4y^2_3\right)M^{\tau}\ .\\     
\end{aligned}
\end{equation}
by which one can see that $u_1$, $u_2$, and $u_3$ are generally irrelevant and flow to zero, and thus the resulting beta functions are
\begin{equation}\label{reduce3}
\begin{aligned}
&\frac{dy_{1}}{dl}=\left(1-M^{\tau}\right)y_{1}\ ,\\
&\frac{dy_{2}}{dl}=y_{2}\ ,\\
&\frac{dy_{3}}{dl}=\left(1-M^{\tau}\right)y_{3}\ ,\\
&\frac{dM^{\sigma}}{dl}=\left(\epsilon -4y^2_1-4y^2_2\right)M^{\sigma}\ ,\\
&\frac{dM^{\tau}}{dl}=\left(\epsilon -4y^2_1-4y^2_3\right)M^{\tau}\ .\\    
\end{aligned}
\end{equation}

\begin{figure}[ht]
   \centering
   \includegraphics[scale=0.43]{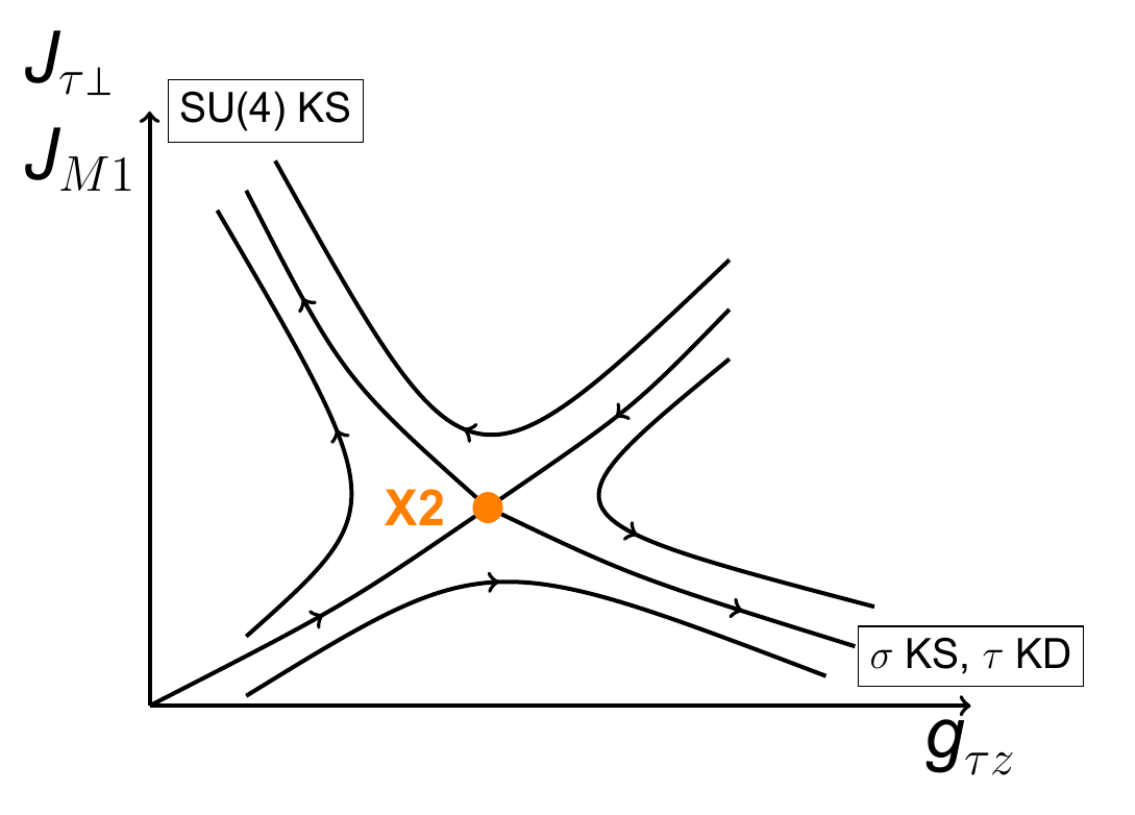}
   \caption{   \label{fig:su4RGre3}   
   {\bf A separate renormalization-group flow.}
   The RG flow diagram of the reduced beta functions (Eq.\,\ref{reduce4}).} 
\end{figure}

From the reduced beta functions (Eq.\,\ref{reduce3}), one can immediately conclude that $y_2\rightarrow \infty$, and thus $M^{\sigma}$ is indeed irrelevant around $M^{\sigma}\rightarrow 0$. The final reduced beta functions are
\begin{equation}\label{reduce4}
\begin{aligned}
&\frac{dy_{1}}{dl}=\left(1-M^{\tau}\right)y_{1}\ ,\\
&\frac{dy_{3}}{dl}=\left(1-M^{\tau}\right)y_{3}\ ,\\
&\frac{dM^{\tau}}{dl}=\left(\epsilon -4y^2_1-4y^2_3\right)M^{\tau}\ .\\
\end{aligned}
\end{equation}
from which one can identify a critical line $\left(y^*_1, y^*_3,M^{\tau *}\right)=\left(a,\frac{\sqrt{\epsilon-4a^2}}{2},1\right)$ where $a$ is a constant, which has one relevant direction with the associated scaling dimension $\sqrt{2\epsilon}$ and separates the spin and orbital KS phase from the spin KS phase and corresponds to the critical point \textbf{X1} in Fig.\,\ref{fig:su4RGresultmain}. The RG flow diagram of reduced beta functions (Eq.\,\ref{reduce4}) 
is plotted in Fig.\,\ref{fig:su4RGre3}. By a parallel analysis, the transition between the spin and orbital KS phase and the orbital KS phase can also be established.

\noindent
\subsection*{E.~RG analysis with a small cross-product term $g_m$}

In this section, we aim to study the stability of the phase diagram 
Fig.\,\ref{fig:su4RGre}(a)
under a small cross-product term $g_m\left(\sigma_z\otimes\tau_{z}\right)\phi_{m}$ through RG analysis. In other words, we derive the beta functions of the Bose-Fermi-Kondo model $H_{BFK}$ where the coupling with the bosonic bath is modified as:
\begin{equation}\label{bkcouplegm}
H_{\rm{BK}}=g_{\sigma z} \sigma_{z}
\phi_{\sigma z} + g_{\tau z} \tau_{z} \phi_{\tau z}+g_m\left(\sigma_z\otimes\tau_{z}\right)\phi_{m}\ .
\end{equation}

After mapping the model into a Coulomb-gas type action, one can derive the beta functions:
\begin{equation}
\begin{aligned}
&\frac{dy_{1}}{dl}=\left(1-K_{1}-M^{\sigma}-M^{\tau}\right)y_{1}+2y_2y_3\ ,\\
&\frac{dy_{2}}{dl}=\left(1-K_{2}-M^{\sigma}-M^{m}\right)y_{2}+2y_1y_3\ ,\\
&\frac{dy_{3}}{dl}=\left(1-K_{3}-M^{\tau}-M^{m}\right)y_{3}+2y_1y_2\ ,\\
&\frac{dK_{1}}{dl}=-2y^2_1\left(2K_1\right)-2y^2_2\left(K_1+K_2-K_3\right)-2y^2_3\left(K_1+K_3-K_2\right)\ ,\\
&\frac{dK_{2}}{dl}=-2y^2_1\left(K_2+K_1-K_3\right)-2y^2_2\left(2K_2\right)-2y^2_3\left(K_2+K_3-K_1\right)\ ,\\  
&\frac{dK_{3}}{dl}=-2y^2_1\left(K_3+K_1-K_2\right)-2y^2_2\left(K_3+K_2-K_1\right)-2y^2_3\left(2K_3\right)\ ,\\      
&\frac{dM^{\sigma}}{dl}=\left(\epsilon -4y^2_1-4y^2_2\right)M^{\sigma}\ ,\\  
&\frac{dM^{\tau}}{dl}=\left(\epsilon -4y^2_1-4y^2_3\right)M^{\tau}\ ,\\     
&\frac{dM^{m}}{dl}=\left(\epsilon -4y^2_2-4y^2_3\right)M^{m}\ .\\  
\end{aligned}
\end{equation}
where $M^m=\Gamma\left(\epsilon\right)g^2_m$.

Since our purpose is only to study the stability of the phase diagram 
Fig.\,\ref{fig:su4RGre}(a)
under small $g_m$, we only need to check the beta function: 
\begin{equation}
\begin{aligned}
&\frac{dM^{m}}{dl}=\left(\epsilon -4y^2_2-4y^2_3\right)M^{m}\\  
\end{aligned}
\end{equation}
by which one can see only the spin and orbital KD fixed point \textbf{G} is unstable against a small $M^m$(and thus $g_m$), while the KS fixed point \textbf{K3}, spin or orbital KS fixed points \textbf{K2} and \textbf{K1}, the generic critical points \textbf{F1}, \textbf{F2}, and the multi-critical point \textbf{R1}, are stable against a weak coupling constant $g_m$.

As a result, the structure of the  the phase diagram 
Fig.\,\ref{fig:su4RGre}(a)
remains unchanged, except now the spin and orbital KD phase correspond to the fixed point:
\[
\textbf{G'}:\;
y_1=y_2=y_3=0,\: M^{\sigma}\rightarrow \infty,\: M^{\tau}\rightarrow \infty,\: M^{m}\rightarrow \infty
\]
instead of \textbf{G}.

\iffalse
\end{widetext}
\fi


\begin{thebibliography}{10}
\bibitem{ignore} Here we ignore a unstable fixed point $\left(y^*_1,y^*,M^*\right)=\left(\frac{\sqrt{\epsilon}}{2},0,\frac{1}{2}\right)$;
because the RG trajactories around it can flow toward the generic critical point  $\left(0,\frac{\sqrt{\epsilon}}{2},1\right)$, it does not influence the structure of the RG flow diagram (Fig.\,\ref{fig:su4RGre}).
\end{thebibliography}
\end{document}